\documentclass[8.5pt,twoside,twocolumn]{article}
\usepackage[super,sort&compress,comma]{natbib} 
\usepackage{mhchem}
 \usepackage{times}
\usepackage{sectsty}
\usepackage{balance} 

\usepackage{graphicx} 
\usepackage{lastpage}
\usepackage[format=plain,justification=raggedright,singlelinecheck=false,font=small,labelfont=bf,labelsep=space]{caption} 
\usepackage{fancyhdr}
\usepackage{graphicx}
\usepackage{amsfonts}
\usepackage{amsmath}
\usepackage{mathtools}
\usepackage{float}
\usepackage{epsfig}
\usepackage{color}
\usepackage{gensymb}
\usepackage{multirow}
\usepackage[mathscr]{euscript}
\usepackage{pifont}

\newcommand{\e}[1]{\ensuremath{\times 10^{#1}}}
\newcommand{\ee}[1]{\ensuremath{10^{#1}}}

\begin{document}

\thispagestyle{plain}
\fancypagestyle{plain}{
\fancyhead[L]{\includegraphics[height=8pt]{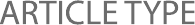}}
\fancyhead[C]{\hspace{-1cm}\includegraphics[height=20pt]{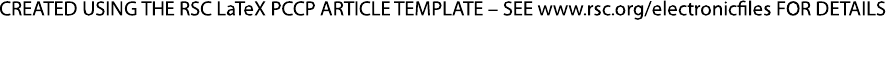}}
\fancyhead[R]{\includegraphics[height=10pt]{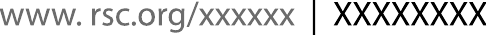}\vspace{-0.2cm}}
\renewcommand{\headrulewidth}{1pt}}
\renewcommand{\thefootnote}{\fnsymbol{footnote}}
\renewcommand\footnoterule{\vspace*{1pt}%
\hrule width 3.4in height 0.4pt \vspace*{5pt}} 
\setcounter{secnumdepth}{5}

\makeatletter 
\def\subsubsection{\@startsection{subsubsection}{3}{10pt}{-1.25ex plus -1ex minus -.1ex}{0ex plus 0ex}{\normalsize\bf}} 
\def\paragraph{\@startsection{paragraph}{4}{10pt}{-1.25ex plus -1ex minus -.1ex}{0ex plus 0ex}{\normalsize\textit}} 
\renewcommand\@biblabel[1]{#1}            
\renewcommand\@makefntext[1]%
{\noindent\makebox[0pt][r]{\@thefnmark\,}#1}
\makeatother 
\renewcommand{\figurename}{\small{Fig.}~}
\sectionfont{\large}
\subsectionfont{\normalsize} 

\fancyfoot{}
\fancyfoot[LO,RE]{\vspace{-7pt}\includegraphics[height=9pt]{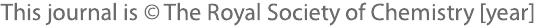}}
\fancyfoot[CO]{\vspace{-7.2pt}\hspace{12.2cm}\includegraphics{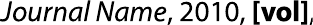}}
\fancyfoot[CE]{\vspace{-7.5pt}\hspace{-13.5cm}\includegraphics{RF-eps-converted-to}}
\fancyfoot[RO]{\footnotesize{\sffamily{1--\pageref{LastPage} ~\textbar  \hspace{2pt}\thepage}}}
\fancyfoot[LE]{\footnotesize{\sffamily{\thepage~\textbar\hspace{3.45cm} 1--\pageref{LastPage}}}}
\fancyhead{}
\renewcommand{\headrulewidth}{1pt} 
\renewcommand{\footrulewidth}{1pt}
\setlength{\arrayrulewidth}{1pt}
\setlength{\columnsep}{6.5mm}
\setlength\bibsep{1pt}

\twocolumn[
  \begin{@twocolumnfalse}
\noindent\LARGE{\textbf{Can exotic disordered ``stealthy'' particle configurations tolerate arbitrarily large holes?}}
\vspace{0.6cm}

\noindent\large{\textbf{G. Zhang,\textit{$^{a}$} F. H. Stillinger,\textit{$^{a}$} and
S. Torquato$^{\ast}$\textit{$^{a, b}$}}}\vspace{0.5cm}

\noindent\textit{\small{\textbf{Received Xth XXXXXXXXXX 20XX, Accepted Xth XXXXXXXXX 20XX\newline
First published on the web Xth XXXXXXXXXX 200X}}}

\noindent \textbf{\small{DOI: 10.1039/b000000x}}
\vspace{0.6cm}

\noindent \normalsize{The probability of finding a spherical cavity or ``hole" of
arbitrarily large size in typical disordered many-particle
systems in the infinite-size limit
(e.g., equilibrium liquid states)
is non-zero. Such ``hole" statistics
are intimately linked to the 
physical properties of the system.
Disordered ``stealthy' many-particle configurations in $d$-dimensional Euclidean space $\mathbb{R}^d$ are exotic amorphous states of matter that lie between a liquid and crystal that prohibit single-scattering events for a range of wave vectors and possess no Bragg peaks [Torquato {\it et al.}, Phys. Rev. X, 2015, {\bf 5}, 021020].
In this paper, we provide strong numerical evidence that disordered stealthy configurations across
the first three space dimensions cannot
tolerate arbitrarily large holes in the infinite-system-size limit,
{\it i.e.}, the hole probability has compact support. This structural ``rigidity" property apparently endows disordered stealthy systems with novel thermodynamic
and physical properties, including 
desirable band-gap, optical and transport characteristics. We also determine
the maximum hole size that any
stealthy system can possess across the first three space dimensions.
}
\vspace{0.5cm}
 \end{@twocolumnfalse}
  ]

\section{Introduction}

Statistical-mechanical studies of disordered many-particle systems often focus on quantifying various statistics of particle locations. This includes n-body correlation functions,\cite{yarnell1973structure, ortiz1994correlation, chandler1987introduction, filipponi1990three}  the structure factor,\cite{yarnell1973structure, chandler1987introduction, ortiz1994correlation}  nearest-neighbor probability distributions,\cite{torquato1990nearest, torquato1995nearest}  and various statistics of the Voronoi cells.\cite{starr2002we, senthil2005voronoi, hentschel2007statistical, slotterback2008correlation, schroder2011minkowski, ma2015tuning}  However, rather than considering the particles themselves, it has been suggested that the space between the particles may be even more fundamental and contain greater statistical-geometrical information.\cite{torquato2001random, torquato2010reformulation}   
A major focus of this paper is the study of a particular property of the void space between point particles in disordered ``stealthy'' systems,\cite{uche2004constraints, batten2008classical, torquato2015ensemble, zhang2015ground, zhang2015ground2} which are disordered many-particle
configurations that anomalously suppress large-scale density fluctuations, endowing them with unique physical properties.
\cite{batten2009novel, florescu2009designer, florescu2013optical, man2013isotropic, leseur2016high, zhang2016transport}  The specific question that we investigate
is whether disordered stealthy systems can contain arbitrarily large holes.
 Here we define a ``hole'' as a spherical region of a certain radius that is empty of particle centers. 
 It is noteworthy that this hole statistic plays a central role in the ``quantizer'' and ``covering'' problems that arise in discrete geometry.\cite{conway2013sphere, torquato2010reformulation}

\begin{figure}
\includegraphics[width=0.49\textwidth]{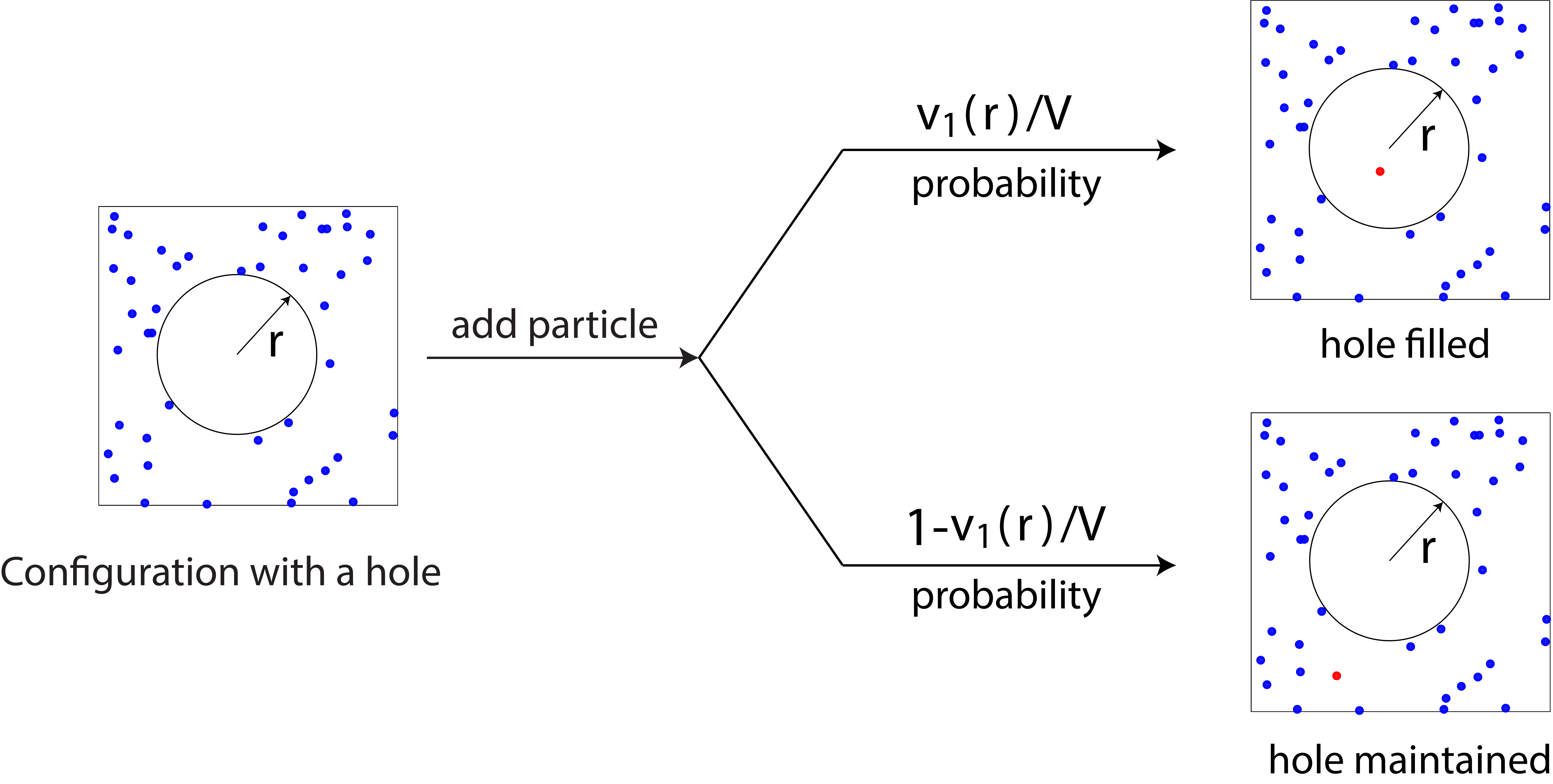}
\caption{In a Poisson point process, particle locations are random and uncorrelated. If there is a hole of volume $v_1(r)$ in a configuration of volume $V$, then when one adds another particle (marked red), the probability that this hole remains empty is $1-v_1(r)/V$. Thus, if there is a total of $N$ particles, the overall probability that such a sphere remains empty is $[1-v_1(r)/V]^N\approx \exp[-(N/V) v_1(r)]=\exp[-\rho v_1(r)]$.}
\label{fig:IdealGas_Ev}
\end{figure}

Given a general many-particle system in $d$-dimensional Euclidean space $\mathbb{R}^d$, can one find arbitrarily large holes?
For disordered systems, the answer to this question is often ``yes.'' Consider the void-exclusion probability function, $E_V(r)$, which gives the probability of finding a randomly located spherical cavity of radius $r$ empty of particles.\cite{torquato2001random} If $E_V(r)$ is non-zero for an arbitrarily large $r$, then one can find arbitrarily large holes in the infinite system, even if these are 
very rare events. For example, as explained in Fig.~\ref{fig:IdealGas_Ev}, the void-exclusion probability for a Poisson point process ({\it i.e.}, an ideal gas) at number density $\rho$ is given by \cite{torquato2001random}
\begin{equation}
E_V(r)=\exp[-\rho v_1(r)],
\label{eq:Ev_Poisson}
\end{equation}
where $v_1(r)=\pi^{d/2}r^d/\Gamma(1+d/2)$ is the volume of a $d$-dimensional sphere of radius $r$,\cite{torquato2010reformulation} and $\Gamma(x)$ is the gamma function. Although $E_V$ decays exponentially as $v_1(r)$ increases, it is always positive for any finite $r$. Thus, no matter how large a hole is desired, the rare event of forming such a hole can always be observed in the infinite system. Similarly, $E_V(r)$ is found to be positive for arbitrarily large $r$'s for equilibrium hard-sphere fluid systems across dimensions.\cite{torquato1990nearest} Therefore, they also allow arbitrarily large holes. It is noteworthy that $E_V(r)$ can be expanded as a series involving n-body correlation functions.\cite{torquato1990nearest}  Therefore, $E_V(r)$ requires many-body correlation information to quantify the probability of hole formation.

Even for many-particle systems in which $E_V(r)$ is not exactly known in the large-$r$ limit, there are often strong arguments indicating that holes of arbitrary sizes can occur. For equilibrium systems of particles interacting with some potentials ({\it e.g.}, Lennard-Jones potential) at some positive temperature $T$, the free energy cost of creating a hole, $\Delta F$, often scales as the hole volume and/or hole surface area, and is therefore finite. Thus, the probability of finding a large hole [roughly $\exp(-\Delta F/T)$] is also nonzero.
Moreover, hard-sphere systems in a glassy or crystalline state away from jamming points possess collective motions that can produce arbitrarily large holes in the infinite-system limit.\cite{atkinson2016static}

Besides the aforementioned many-particle systems with unbounded hole sizes, we also know of several systems in which the hole radii are bounded
from above.
A simple class of systems whose hole probability must
have compact support are perfect
crystalline (periodic) many-particle systems. Spheres large enough to encompass entire unit cells always contain particles. Thus, holes of arbitrarily-large radii cannot exist. 
A simple disordered class is saturated random sequential addition (RSA) sphere packings across dimensions. RSA is a time-dependent packing process, in which congruent hard spheres are randomly and sequentially placed into a system without overlap. In the infinite-time
limit, the system becomes saturated, {\it i.e.,} spheres can no longer be added to the packing, and hence holes must be finite in size. By contrast, RSA packings below the saturation density were found to have positive $E_V(r)$ for arbitrarily large $r$,\cite{rintoul1996nearest} and therefore allow for the presence of very large holes.

So far we have seen that although all perfect crystalline many-particle systems prohibit arbitrarily large holes, many disordered many-particle systems allow them. A promising class of amorphous structures that may not tolerate
arbitrarily large holes is
disordered hyperuniform systems.
Such systems have received considerable attention because they anomalously suppress density fluctuations.\cite{batten2009novel, florescu2009designer, florescu2013optical, man2013isotropic, leseur2016high, zhang2016transport} 
Specifically, if one places a spherical window of radius $R$ into a $d$-dimensional many-particle system and counts the number of particles in the window, then the number variance, $\sigma^2(R)$, scales as $R^d$ for large $R$ in typical disordered systems. Any system in which $\sigma^2(R)$ grows slower than $R^d$ is said to be hyperuniform.\cite{torquato2003local} Equivalently, a hyperuniform
many-particle system is one which
the structure factor $S({\bf k})$
tends to zero as the wavenumber
$|{\bf k}|$ tends to zero,\cite{torquato2003local} {\it i.e.,}
\begin{equation}
\lim_{|\mathbf k| \to 0} S(\mathbf k) =0.
\end{equation}
Disordered hyperuniform systems are a good starting point to search for more examples of disordered systems with bounded hole sizes because the formation of large holes might be inconsistent with hyperuniformity, which suppresses large-scale density fluctuations.

However, we know that not all disordered hyperuniform systems prohibit arbitrarily large holes. For example, in a hyperuniform fermionic-point 
process in $d$ spatial dimensions, $E_V(r)$ scales as $\exp(-cr^{d+1})$ (where $c$ is a constant) for large $r$.\cite{torquato2008point}
Also, the hyperuniform two-dimensional one-component plasma possesses an $E_V(r)$ that scales as $\exp(-cr^{4})$ for large $r$.\cite{Hough09, ghosh2016gaussian}
Both of these systems thus allow arbitrarily large holes.
Therefore, hyperuniformity alone is not a sufficient condition to guarantee boundedness of the hole size. Nevertheless, different hyperuniform systems have different levels of suppression for large-scale density fluctuations. While any system in which $\lim_{|\mathbf k| \to 0} S(\mathbf k)=0$ is considered hyperuniform, the ``stealthy'' variants of hyperuniform systems have $S(\mathbf k)=0$ in the entire interval $|\mathbf k|\in (0, K]$ for a certain value of $K$. 
Stealthy hyperuniform systems are known to possess many unique physical properties, including negative thermal expansion behavior,\cite{batten2009novel} complete isotropic photonic band gaps comparable in size to those of a photonic crystal,\cite{florescu2009designer, florescu2013optical, man2013isotropic} transparency even at high densities,\cite{leseur2016high} and nearly optimal transport properties.\cite{zhang2016transport}
The behavior of $S(\mathbf k)$ near $\mathbf k=\mathbf 0$ in stealthy systems is identical to that in perfect crystals. Since perfect crystals  prohibit large holes, could stealthy hyperuniform systems also prohibit large holes?

In this paper, we present strong numerical evidence that disordered stealthy systems indeed prohibit arbitrarily large holes.
It is nontrivial to study the existence of large holes not only because formation of large holes is extremely rare, but also because numerical simulations are limited to finite-sized systems and one wants to infer the infinite-volume-limit behaviors. 
With periodic boundary conditions, such systems are always perfect crystals, even if the repeating units may be very large. As we have mentioned, perfect crystals always have bounded hole sizes.
We developed two numerical techniques to overcome these issues to distinguish whether a system can tolerate arbitrarily large holes or not that can be applied to infer 
the maximum hole size
in general disordered systems
(whether they are stealthy or not)
in the infinite-volume limit.
Specifically, we first attempt to determine the maximum size of the holes that naturally emerges in stealthy hyperuniform systems across the first three space dimensions by studying the tail behavior of $E_V(r)$. We find that the tail of $E_V(r)$ for stealthy systems is qualitatively similar to that for crystals and saturated RSA sphere packings, which have finite holes, and is qualitatively different from that for Poisson point processes with unbounded hole sizes.
We then determine
the maximum hole size that any
stealthy system can possess across the first three space dimensions.
To do this, we generate large stealthy systems with largest possible holes by imposing repulsion fields with sizes equal to the desired hole sizes in stealthy systems. We discover that this method can only create holes of certain finite sizes without breaking stealthiness. In stealthy configurations with largest possible holes, particles concentrate in concentric shells around the hole. Analytical studies on this pattern allows us to derive a conjectured upper bound of the hole radius for all stealthy systems.

The rest of the paper is organized as follows: Section~\ref{sec:Definitions} defines stealthy point patterns and two associated parameters, $\chi$ and $K$. Section~\ref{sec:Unbiased} studies maximum hole sizes and the tail behavior of $E_V(r)$ in such systems. Section~\ref{sec:FieldStudy} defines the repulsion field we used to create holes, study the pattern of stealthy systems with such holes, and conjecture an upper bound for the hole radius, in one to three dimensions. Section~\ref{sec:Conclusions} provides concluding remarks and discussions.

\section{Mathematical definitions}
\label{sec:Definitions}

For a single-component system with $N$ particles, located at $\mathbf r^N=\mathbf r_1, \mathbf r_2, \cdots, \mathbf r_N$, in a simulation box of volume $V$ with periodic boundary conditions in a $d$-dimensional Euclidean space $\mathbb R^d$, the static structure factor is defined as $S(\mathbf k)=|\sum_{j=1}^N \exp(-i \mathbf k \cdot \mathbf r_j)|^2/N$, where $i$ is the imaginary unit and $\mathbf k$ is a $d$-dimensional wavevector (which must be integer multiples of the reciprocal lattice vectors of the simulation box).\cite{chandler1987introduction, chaikin2000principles}

As we have explained earlier, a hyperuniform system is defined as one in which the number variance $\sigma^2(R)$ grows more slowly than $R^d$ for large window radius $R$, or a system in which $\lim_{|\mathbf k| \to 0} S(\mathbf k)=0$.
Stealthiness is a stronger condition than hyperuniformity. For some positive $K$, we call a system ``stealthy up to $K$'' if
\begin{equation}
S(\mathbf k) =0 \mbox{ for all $0<|\mathbf k|<K$.}
\label{eq:StealthyDefinition}
\end{equation}

In this paper, we define the following potential energy  function to be a ``stealthy potential'' \footnote{This definition of $\Phi_s$ actually differs from previous definitions of ``stealthy potentials'' \cite{zhang2015ground} by a constant, which has no effect on the configurational behavior of the system.}
\begin{equation}
\Phi_s(K; {\mathbf r}^N) =\frac{N}{2V} \sum_{0<|\mathbf k|<K} {\tilde v}({\bf k})S(\mathbf k),
\label{eq:StealthyPotential}
\end{equation}
where ${\tilde v}({\bf k})$ is a positive function of $\mathbf k$. For present purposes, we choose ${\tilde v}({\bf k})=1$ for simplicity.
Because $S(\mathbf k)$ is by definition always non-negative, the ground-state energy of this potential is zero, and the set of the ground states is equal to the set of configurations stealthy up to $K$. 

Only half of the constraints in Eq.~(\ref{eq:StealthyDefinition}) are independent. This is because by definition, $S(\mathbf k)=S(-\mathbf k)$. Let the number of independent constraints be $M$, so the parameter
\begin{equation}
\chi=\frac{M}{d(N-1)}
\end{equation}
quantifies the fraction of degrees of freedom that is constrained. Because $\chi$ is proportional to $M$, it is also proportional to $v_1(K)$, the volume of a $d$-dimensional sphere of radius $K$. Indeed, we have previously found \cite{torquato2015ensemble}
\begin{equation}
\rho\chi=\frac{v_1(K)}{2d(2\pi)^d}.
\label{eq:rhochi}
\end{equation}

It was found that for $\chi<0.5$, the ground states of stealthy potentials are uncountably infinitely degenerate, and possess no long-range order.\cite{zhang2015ground} As $\chi$ increases beyond 0.5, the ground states are still uncountably infinitely degenerate, but develop long-range translational and orientational order.\cite{zhang2015ground2}
As $\chi$ increases further, these ground states eventually undergo phase transitions into the integer lattice, the triangular lattice, and the BCC lattice in one, two, and three dimensions, respectively.\cite{torquato2015ensemble} 
In this paper, we want to study hole sizes of disordered stealthy systems, and will therefore focus on the $\chi<0.5$ range.
Because ground states of the stealthy potentials are uncountably infinitely degenerate, one can have different ways to sample the ground states, which assign different weights to different parts of the ground state manifold. We have previously focused on the zero-temperature limit of the canonical ensemble ({\it i.e.}, define the probability measure $P({\mathbf r}^N) \propto \exp[-\Phi_s(K; {\mathbf r}^N)/k_BT]$, where $k_B$ is the Boltzmann constant and $T$ is the temperature, and then take the $T \to 0$ limit). 
However, in this paper, we will also assign different weights to bias toward configurations with large holes.

\section{Hole Probability and Maximum Hole Size in Unbiased Stealthy Systems}
\label{sec:Unbiased}

If an upper bound on the hole sizes exists, how should it depend on $K$ and $\chi$? The $K$ dependence can be easily ascertained from a scaling argument: If there exists a configuration with hole size $R$ that is stealthy up to $K$, then by rescaling the real-space configuration by a factor $\alpha$, one can create another configuration with hole size $R\alpha$, stealthy up to $K/\alpha$. Therefore, the maximum hole radius, $R_c$, must be inversely proportional to $K$.
Therefore, we henceforth study the dimensionless hole size, $R_cK$, rather than $R_c$ itself.

A different argument can shed light on the dependence of the hole size on $\chi$. 
A superposition of multiple configurations, each stealthy up to a certain $K$, is also stealthy up to the same $K$.\cite{torquato2015ensemble} Therefore, if there exist $n$ configurations, each with a hole of radius $R$ that is stealthy up to $K$, then one could superpose them with hole centers aligned to create another configuration with the same hole radius $R$ and $K$. However, since the number of particles increases by a factor of $n$, $\chi$ decreases by a factor of $n$. Therefore, if there exists a configuration of a certain hole size and $K$ at some $\chi$ value, then there exists a configuration of the same hole size and $K$ at arbitrarily small $\chi$ values. In other words, $R_cK$ as a function of $\chi$ must achieve the global maximum in the $\chi \to 0^+$ limit.

\begin{figure}
\includegraphics[width=0.49\textwidth]{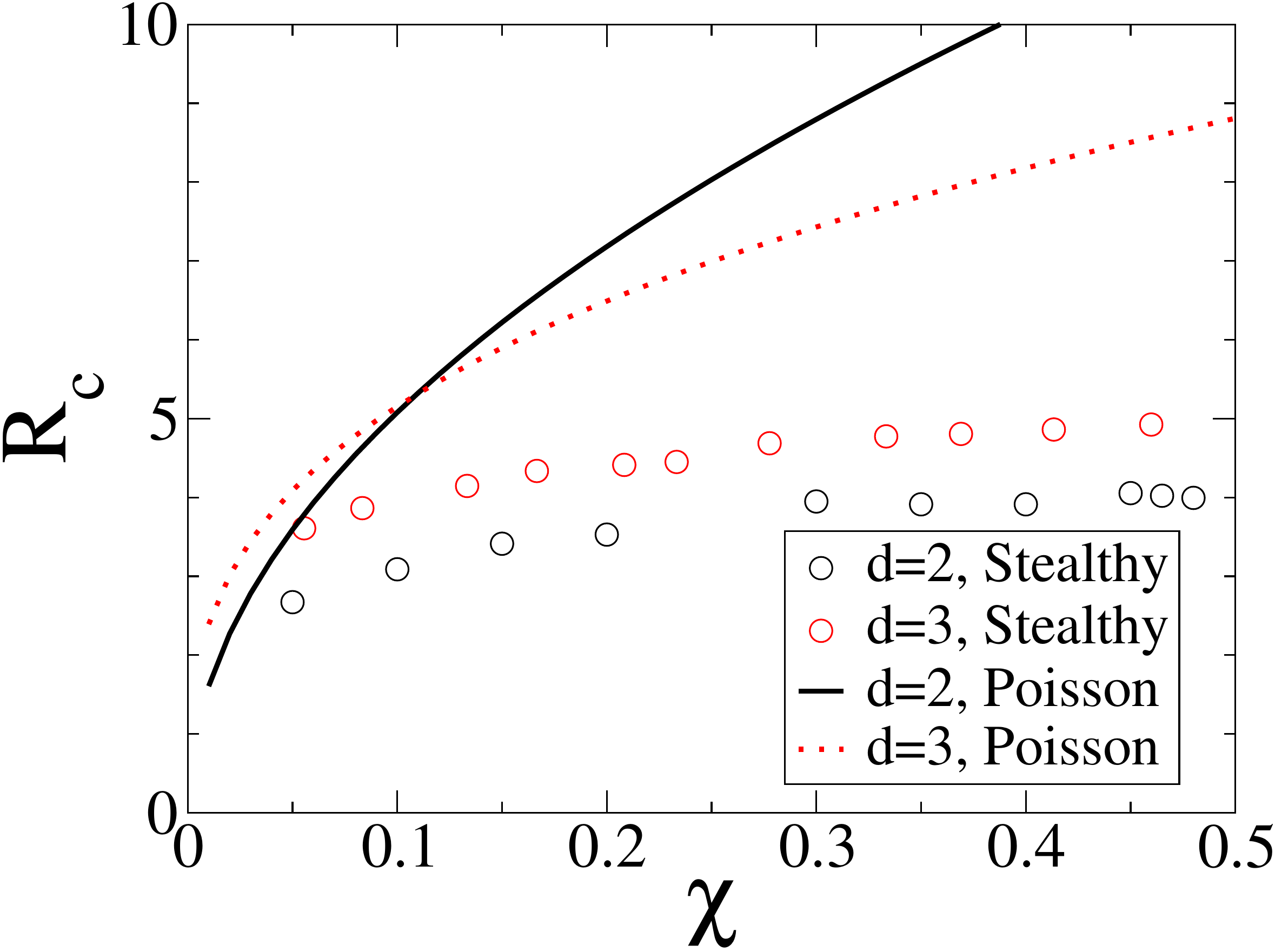}
\caption{Maximally observed $R_c$ in 20,000 entropically favored stealthy ground states, rescaled to unity $K$. The number of particles per configuration, $N$, depends on $\chi$ and space dimensions but is always between 421 and 751 and is given in Ref.~\citenum{zhang2016transport}. The same quantity for Poisson point processes (ideal gas) at the same density is also plotted for comparison.}
\label{fig:Stealthy_Rc}
\end{figure}

With these preliminary analytical results in mind, let us examine the numerical results from unbiased ground states of stealthy potentials ({\it i.e.}, $T=0$ limit of the canonical ensemble).
We have previously generated such ground states in two and three dimensions for various $\chi$ values by performing low-temperature ($k_BT=2\e{-6}$ in 2D and $k_BT=\ee{-6}$ in 3D) molecular dynamics simulations, periodically taking snapshots, and then minimizing the energy starting from each snapshot; see Ref.~\citenum{zhang2016transport} for more details. For each $\chi$, we generated 20,000 configurations. The number of particles, $N$, is always between 421 and 751 and is detailed in Ref.~\citenum{zhang2016transport}. 
For each configuration, we rescaled it to unity $K$ and performed a Voronoi tessellation and found out the largest distance between each Voronoi vertex and its neighbor particles. This distance is the maximum hole size for any particular configuration. We then determined the maximum hole size among all 20,000 configurations and plotted them as a function of $\chi$ in Fig.~\ref{fig:Stealthy_Rc}. 
For a comparison, we also present the same quantity for Poisson point processes at the same conditions, derived in Appendix~\ref{app:PoissonRc}. As Eq.~(\ref{eq:rhochi}) shows, with $K$ fixed to unity, $\rho$ is inversely proportional to $\chi$. Thus, it is not surprising that $R_c$ for Poisson processes increases as $\chi$ increases.  
In unbiased stealthy ground states, however, $R_c$ weakly increases with increasing $\chi$ and saturates at some constant value, suggesting that $R_c$ is bounded for stealthy ground states with fixed $K$.
The critical radius $R_c$ decreases slightly as $\chi$ tends to zero
because unbiased stealthy ground states become less ordered. Therefore, although large hole formation is still possible, its probability decreases. When this probability is too low, it becomes computationally more difficult to find such a large hole with only 20,000 configurations. 

\begin{figure}
\includegraphics[width=0.49\textwidth]{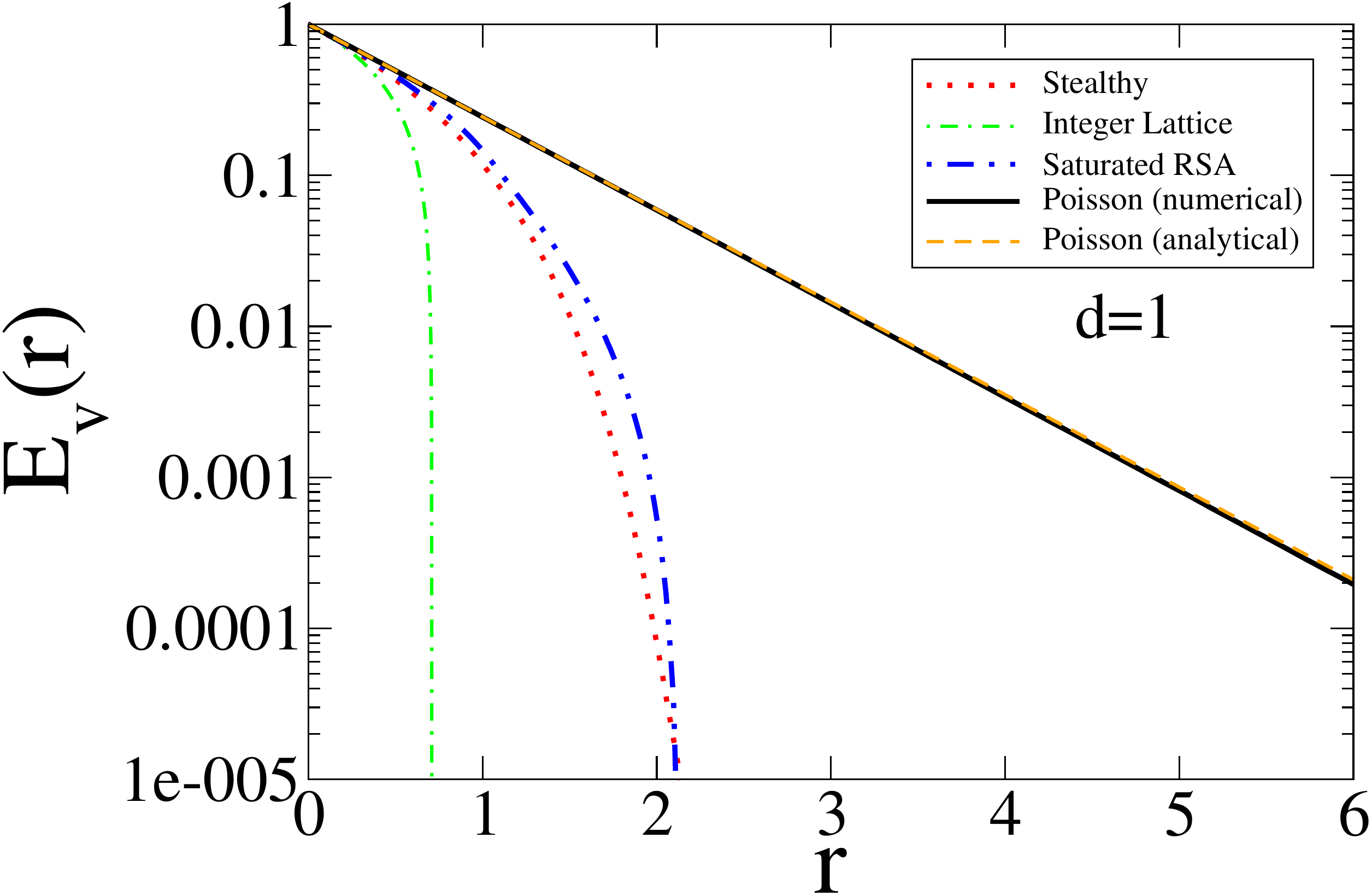}
\includegraphics[width=0.49\textwidth]{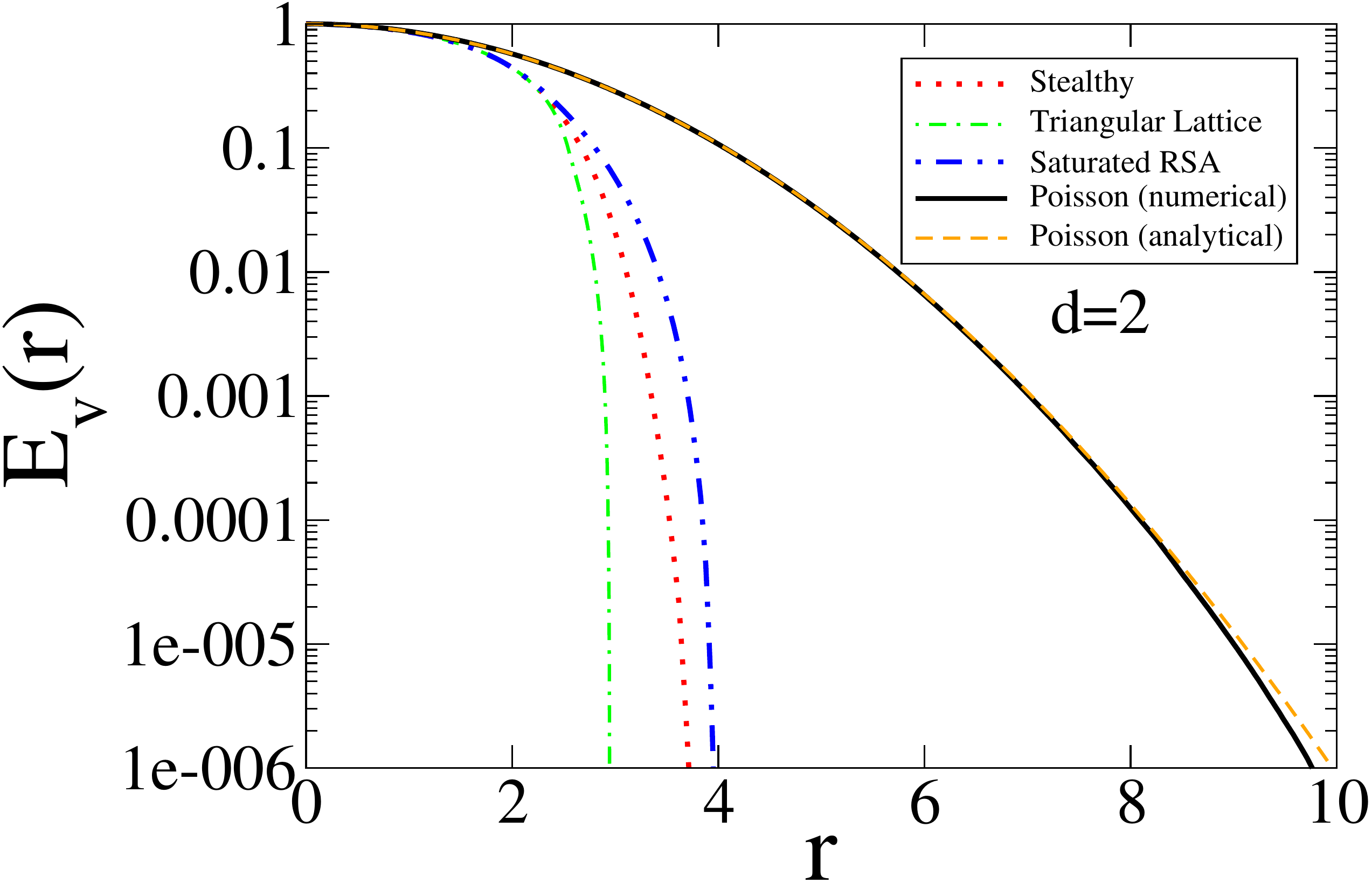}
\includegraphics[width=0.49\textwidth]{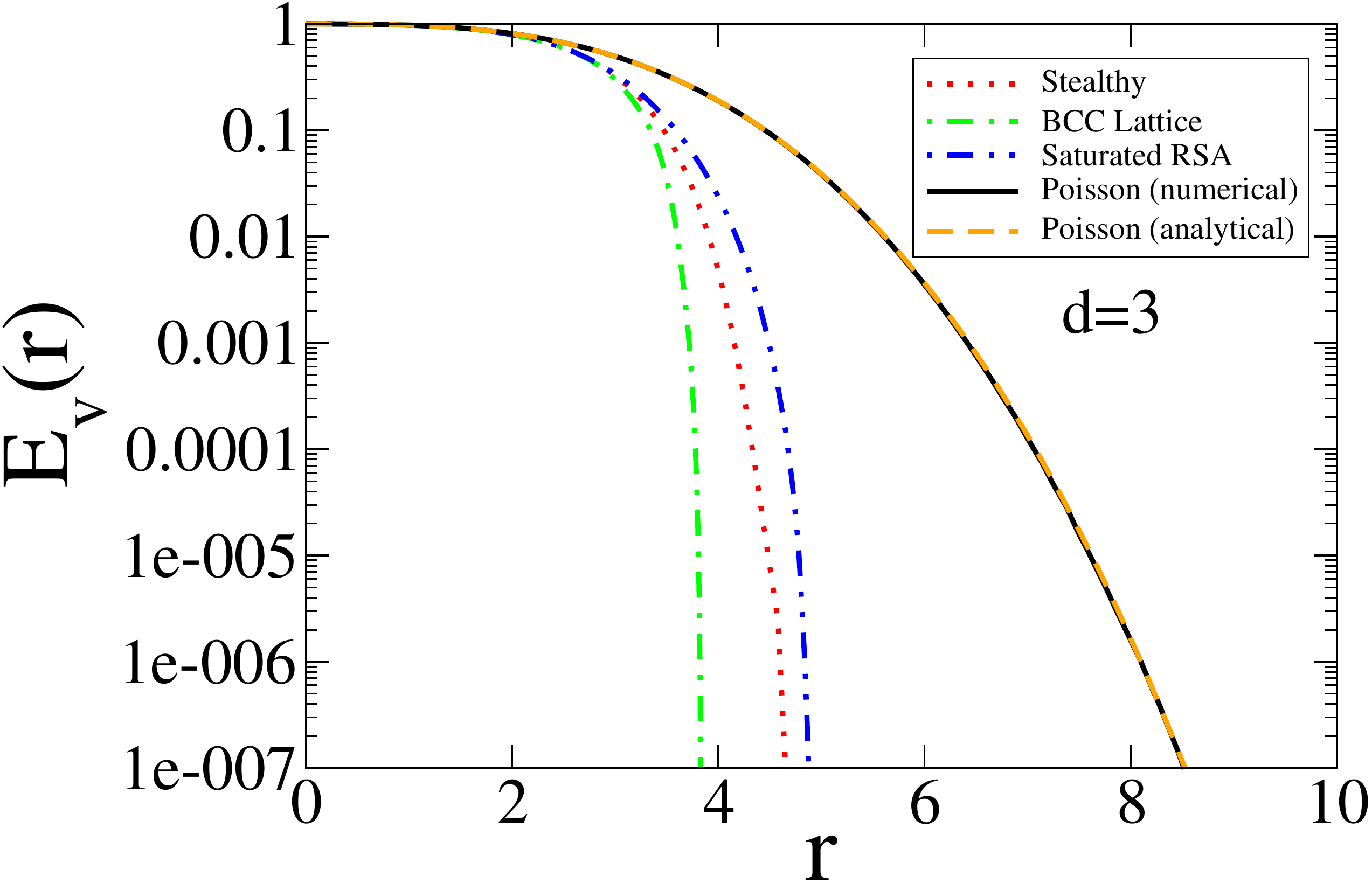}
\caption{Numerically computed $E_V(r)$ for (top) a stealthy system at $\chi=0.45$ in 1D, (middle) a stealthy system at $\chi=0.45$ in 2D, and (bottom) a stealthy system at $\chi=0.46$ in 3D. For comparison, we also present $E_V(r)$ of perfect crystals (integer, triangular, and BCC lattices\cite{torquato2010reformulation}), saturated RSA packings, and Poisson point processes at the same number density across the first three
space dimension. 
For Poisson point processes, we present both numerically found $E_V(r)$ and exact analytical predictions for $E_V(r)$. The  excellent agreement between these numerical and exact results is a testament
to the numerical precision
of our calculations.
}
\label{fig:Stealthy_Ev}
\end{figure}

Examining the large-$r$ tail behavior of $E_V(r)$ suggests strongly that $R_c$ is finite in stealthy systems. As we have explained in Sec.~I, if the hole size is bounded, $E_V(r)$ for some value of $r$ must be identically zero, instead of being exponentially small. In Fig.~\ref{fig:Stealthy_Ev}, we closely examine the tails of $E_V(r)$ of stealthy systems in the first three space dimensions in a semi-log scale. 
As we showed earlier, numerically found $R_c$ suffer from greater sampling errors if $\chi$ is too small. Thus, to study the tail behavior of $E_V(r)$, we choose sufficiently large $\chi$ values (0.45-0.46) in Fig.~\ref{fig:Stealthy_Ev}. Nevertheless, we will show in the next section that smaller $\chi$ values do not result in any qualitative difference.
For purposes of comparison, we compare our results for stealthy systems to $E_V(r)$ for systems in which we know that the holes must be finite in size, namely, lattices
in which $E_V(r)$ is given exactly\cite{torquato2010reformulation} and saturated RSA sphere packings; and contrast our results to Poisson point processes, in which hole sizes are unbounded.
As Fig.~\ref{fig:Stealthy_Ev} shows, the tail behavior of stealthy systems resembles that of crystalline structures and saturated RSA packings.
For each of these systems, the logarithm of $E_V(r)$ must decay to its bounded cut-off value of $R_c$
with an infinite slope at which $E_V(R_c)=0$, which may be regarded to be singularity. However, these figures necessarily present $E_V(r)$ above certain positive lower limits and hence only nearly-infinite slopes are apparent.
By contrast, Poisson point processes and equilibrium hard-sphere fluids (not shown in the figure), which have unbounded $R_c$'s, possess $\log [E_V(r)]$'s that comparatively have very small slopes on the scale of the figures, without any singularity.
Note that although $E_V(r)$ of RSA packings have been studied before,\cite{rintoul1996nearest, zhang2013precise} this is the first study that focuses on its tail behavior. 

It is noteworthy that the three lattice structures we chose (integer, triangular, and BCC lattice) are the optimal solutions of the covering and quantizer problems \cite{conway2013sphere} in their respective dimensions. In a specific dimension and density, the covering problem asks for the configuration with the smallest cutoff in $E_V(r)$ ({\it i.e.,} the smallest $R_c$), while the quantizer problem asks for the configuration that minimizes the so-called ``quantizer error,'' defined as\cite{torquato2010reformulation}
\begin{equation}
\mathcal{G}=\frac{2}{d}\int_0^\infty r E_V(r) dr.
\end{equation}
As Fig.~\ref{fig:Stealthy_Ev} shows, in two and three dimensions, $E_V(r)$ of stealthy systems at $\chi=0.45-0.46$ is quite close to $E_V(r)$ of the triangular and BCC lattices. Therefore, stealthy ground states at high $\chi$ values should provide nearly optimal solutions to these two problems.

\section{Stealthy configurations with largest possible holes}
\label{sec:FieldStudy}

In the previous section we studied the largest holes naturally occurring in unbiased disordered ground states of stealthy potentials. 
In this section, we study the maximum hole sizes
consistent with stealthiness. To do so, we impose a radial exclusion field at the center of the simulation box to bias the configuration toward ones with largest holes. We combine the stealthy potential with such an exclusion field, and try to find the ground state of the system. We then study the patterns of the resulting ground states.

\subsection{Simulation details}
\label{sec:Simulation}

To bias toward configurations with large holes, we let the total potential energy be a sum of the stealthy potential contribution and the exclusion field contribution:
\begin{equation}
\Phi(\mathbf r^N)=\Phi_{s}(K; \mathbf r^N)+\Phi_{ex}(R_f; \mathbf r^N),
\end{equation}
where $\Phi_{s}(K; \mathbf r^N)$ is the stealthy potential given in Eq.~(\ref{eq:StealthyPotential}), and $\Phi_{ex}(R_f; \mathbf r^N)$ is the exclusion-field contribution, given by
\begin{equation}
\Phi_{ex}(R_f; \mathbf r^N)=\sum_{i} F(R_f; r_{ic}),
\label{eq:field1}
\end{equation}
where $r_{ic}$ is the radial distance from particle $i$ to the center of the simulation box, 
\begin{equation}
F(R_f; r_{ic})=
\begin{cases}
(R_f/r_{ic}-1),& \text{if $r_{ic}<R_f$,} \\
0,& \text{otherwise,}
\end{cases}
\label{eq:field}
\end{equation}
and $R_f$ is the radius of the exclusion field. By varying $R_f$, we can probe the largest possible hole size in a particular system. Before $R_f$ reaches $R_c$ (the upper bound of the hole radius), $\Phi_{ex}$ can be zero. However, once $R_f$ surpasses $R_c$ for a particular system, $\Phi_{ex}$ must be positive.

If we can find a configuration for which $\Phi(\mathbf r^N)=0$, then both $\Phi_{s}(K; \mathbf r^N)$ and $\Phi_{ex}(R_f; \mathbf r^N)$ must be zero, and therefore this configuration is stealthy up to $K$ while simultaneously having a hole radius $R_f$. To test if there are such configurations, we perform energy minimizations using the L-BFGS algorithm,\cite{nocedal1980updating, liu1989limited, nlopt}  starting from many random initial configurations, and finding if the ending $\Phi(\mathbf r^N)$ in any configuration dropped below a strong tolerance of $\ee{-10}$. We consider a certain number, $R_c$, to be the numerically found maximum hole size if a zero-energy configuration is found within $N_{trial}$ energy minimization trials for $R_f=R_c$, but not found for $R_f=R_c+\delta_R$. Here we choose $N_{trial}=100$ and $\delta_R=0.01$. For a two-dimensional system at $\chi=0.10$, and $N=400$, with this choice of $N_{trial}$ and $\delta_R$ we find $R_cK=4.58$; while using $N_{trial}=1000$ and $\delta_R=0.0001$, we find $R_cK=4.5903$. Therefore, our choice of $N_{trial}$ and $\delta_R$ produces $R_cK$ values with approximately $\ee{-2}$ precision. As explained in our previous work,\cite{zhang2015ground} to minimize boundary effects for the stealthy potential, we use a rhombic simulation box with a $60^\circ$ interior angle in 2D and a simulation box in the shape of a fundamental cell of
a body-centered cubic lattice in 3D with periodic boundary conditions. 

\begin{figure}
\includegraphics[width=0.49\textwidth]{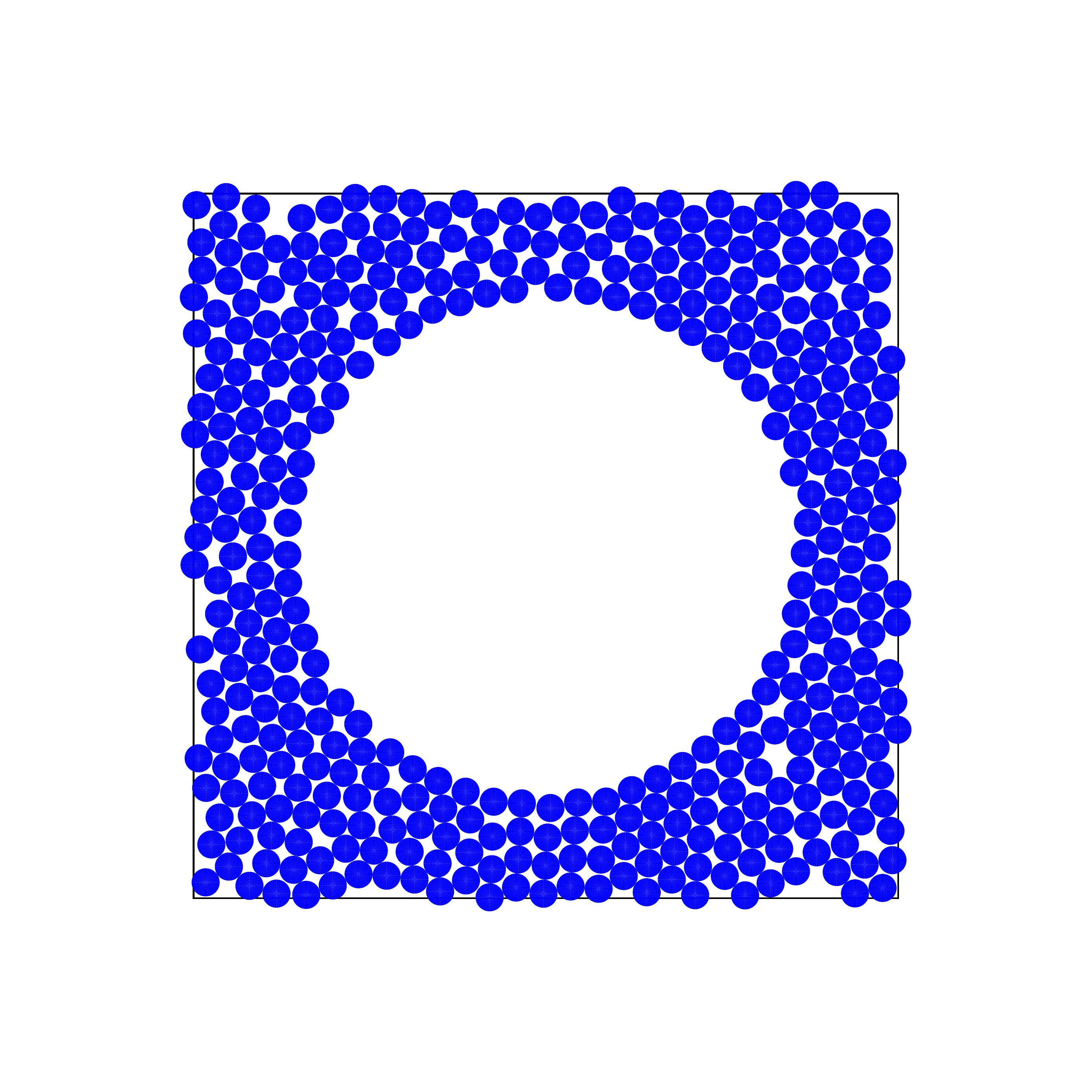}
\caption{A configuration obtained by energy minimization using the potential in Eq.~(\ref{eq:OHern}) and an external field of radius $R_f=9.2$. The simulation box contains $N=400$ particles and has side length $L=25$.}
\label{fig:OHern_Exclusion}
\end{figure}

As a test for this methodology, we combined the exclusion field [Eq.~(\ref{eq:field1})] with following pair potential
\begin{equation}
\Phi_h(\mathbf r^N)=\sum_{i<j} v_2(r_{ij}),
\label{eq:OHern}
\end{equation}
where
\begin{equation}
v_2(r_{ij})=
\begin{cases}
(1-r_{ij})^2,& \text{if if $r_{ij}<1$} \\
0,& \text{otherwise,}
\end{cases}
\label{eq:OHernPair}
\end{equation}
and performed energy minimizations in two dimensions. For this potential to be zero, any pair of particles cannot be closer than distance 1. Therefore, the ground state of this potential corresponds to an equilibrium hard disk system of diameter 1. As we have mentioned in Sec.~I, any such system in the infinite-volume
limit must possess an unbounded hole size. Nevertheless, the formation of very large holes is still very rare and may be difficult to observe if one simply samples unbiased configurations.
We performed our simulation on an $N=400$ system with volume fraction $\eta=0.5$. 
As shown in Fig.~\ref{fig:OHern_Exclusion}, the energy minimization algorithm is capable of creating a hole of of radius $R=9.2$, although the probability of finding such a hole in an unbiased system is extremely small. According to Eq.~(4.21) of Ref.~\citenum{torquato1990nearest}, $E_V(9.2)=4\e{-279}$.
This demonstrates that if the hole size is unbounded in the infinite-system-size limit for some system, this numerical protocol can indeed create very large holes in a finite-size simulation.
Figure~\ref{fig:OHern_Exclusion} also shows that in creating such a large hole, the particles are pushed to each other as closely as possible ({\it i.e.,} up to interparticle contacts). Therefore, even larger holes should be possible if we simulated larger systems at the same volume fraction.

\subsection{One-dimensional study}
\label{sec:OneD}

\begin{figure}
\includegraphics[width=0.49\textwidth]{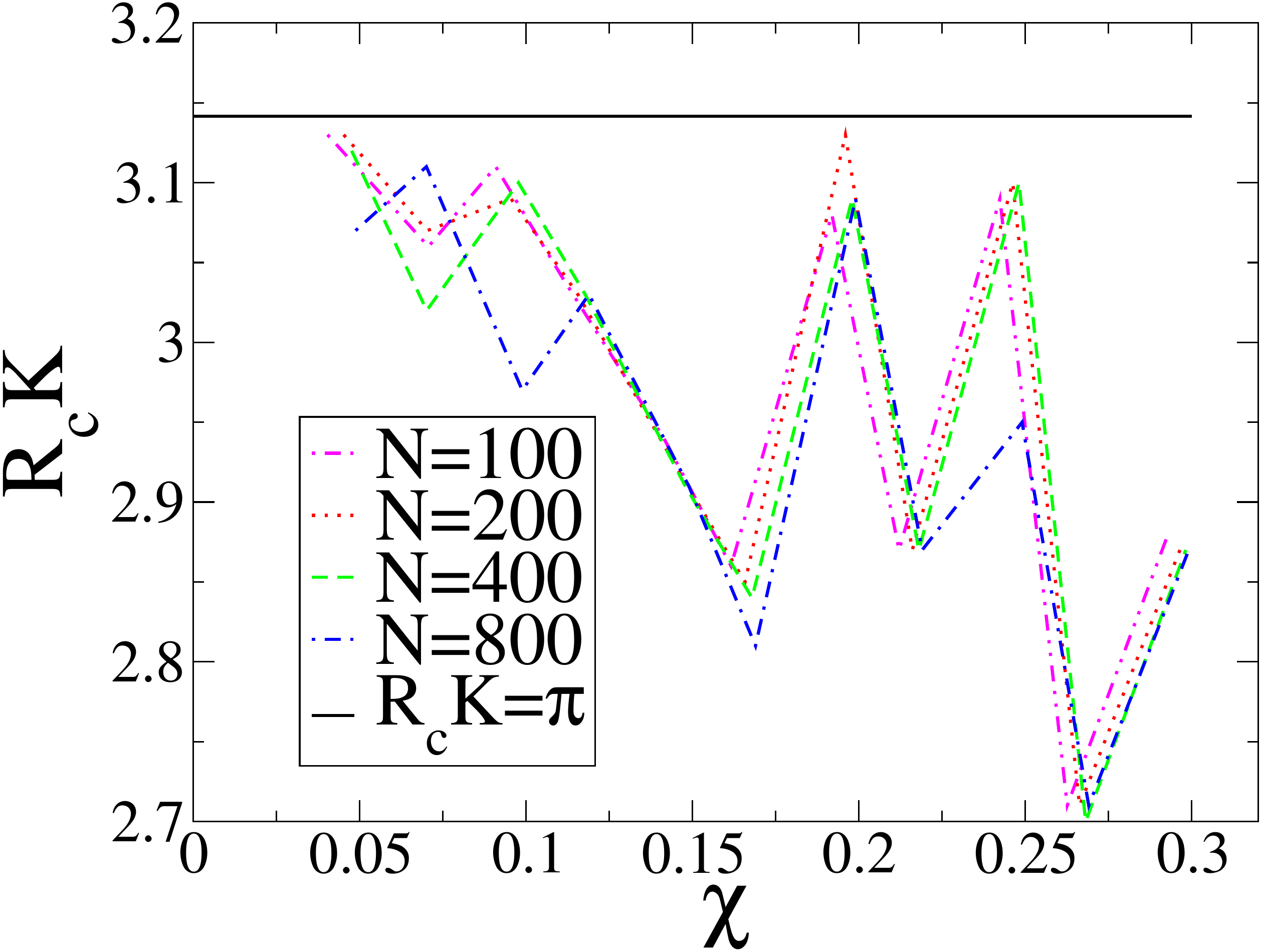}
\caption{Numerically found maximum $R_cK$, as a function of $\chi$, in 1D biased stealthy configurations for various system sizes.}
\label{fig:1D_Rc_chi}
\end{figure}

We first examine $R_cK$ values found by the above-mentioned algorithm in 1D, since this is computationally the easiest dimension to study and will shed light on corresponding results in higher dimensions. Our result for several different $\chi$'s and system sizes are summarized in Fig.~\ref{fig:1D_Rc_chi}. It appears that $R_cK$ as a function of $\chi$ is chaotic and displays no systematic trend. Nevertheless, Fig.~\ref{fig:1D_Rc_chi} does show that $R_cK$ is always close to $\pi$ but never exceeds it. As we will see later, $\pi$ is the upper bound of $R_cK$ in 1D.

\begin{figure}
\includegraphics[width=0.48\textwidth]{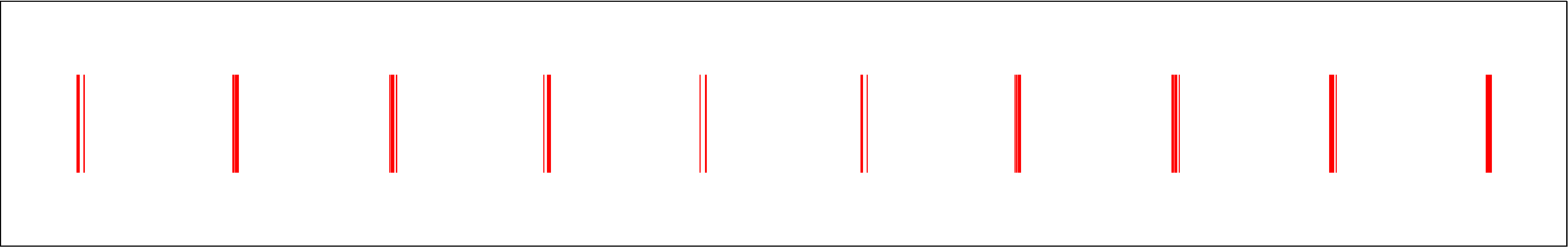}
\caption{A 1D biased stealthy configuration of $N=100$ particles obtained by energy minimization using the stealthy potential of $K=1$ and an external field of radius $R_f=3.1$ at $\chi=0.1$. The particles self-assemble into 10 clusters. Although particles in the same cluster may not be distinguishable from each other here, we have examined the configuration and find that each cluster contains exactly 10 particles.}
\label{fig:1DExclusion}
\end{figure}

Examining stealthy configurations with hole sizes $R_f\approx\pi/K$ reveals a more interesting behavior. Such a configuration is shown in Fig.~\ref{fig:1DExclusion}. At exclusion-field size $R_f=3.1/K$, 100 particles self-assemble into 10 clusters, each containing 10 particles. These clusters then form a one-dimensional integer lattice. 

As we have explained in Sec.~\ref{sec:Unbiased}, a superposition of multiple integer lattices, with hole centers aligned, have the same $R_cK$ as a single integer lattice. It is straightforward to calculate $R_cK$ of an integer lattice: If the distance between neighboring lattice sites is $L$, then the maximum hole radius is $L/2$, and the stealthy range $K$ is equal to the location of the first Bragg peak, $2\pi/L$. Therefore, $R_cK$ of any integer lattice is simply $\pi$. To summarize, the numerically found hole radius is never above $\pi/K$; and superposed integer lattices can indeed achieve hole radius $\pi/K$. Therefore, we expect that $\pi/K$ is an upper bound of the hole size for stealthy 1D structure at any $\chi$.

\subsection{Two- and Three-dimensional studies}
\label{sec:HigherD}

\begin{figure}
\includegraphics[width=0.49\textwidth]{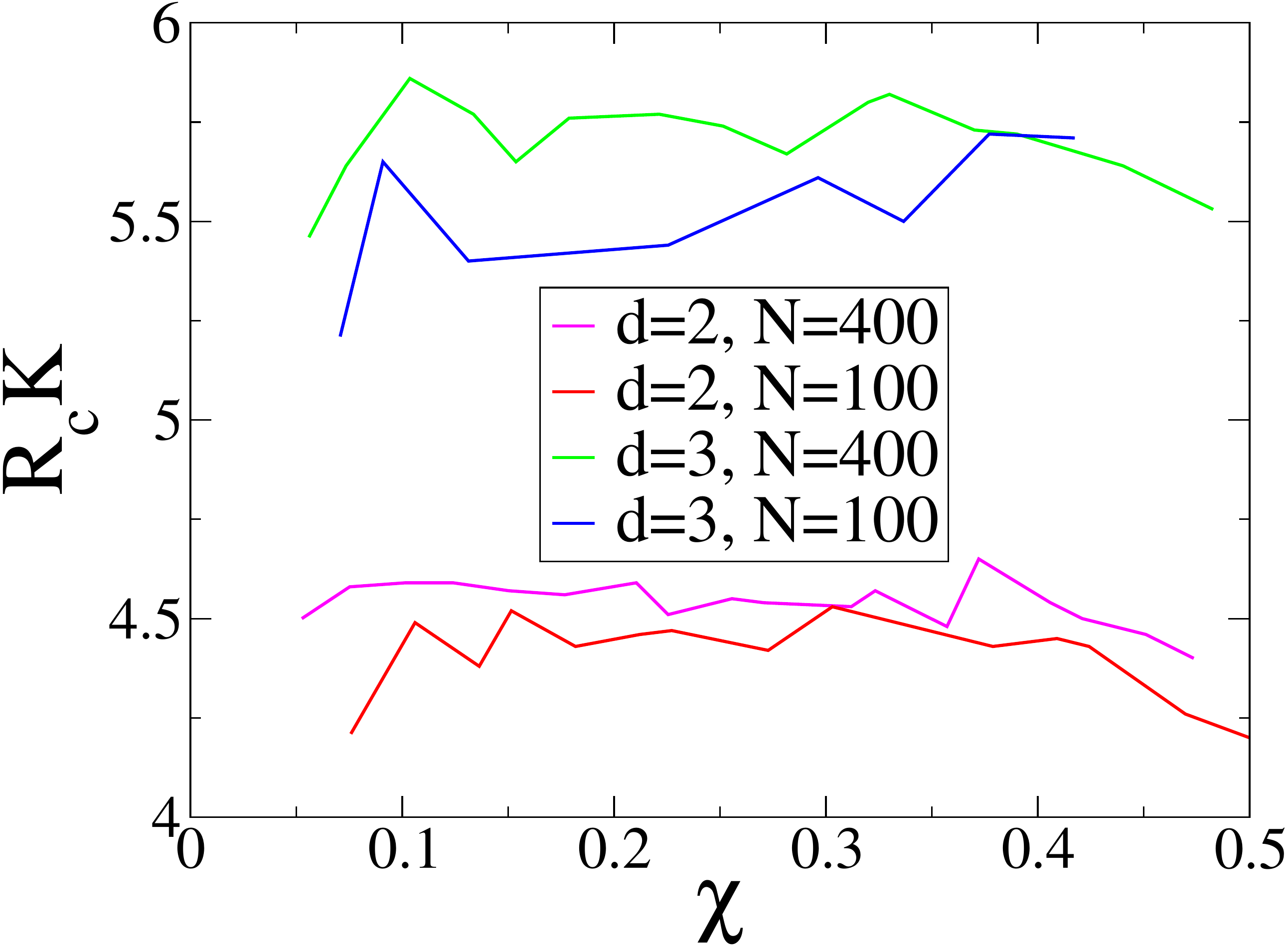}
\caption{Numerically obtained maximum $R_cK$, as a function of $\chi$, in 2D and 3D biased stealthy configurations. }
\label{fig:NumericalRc_K}
\end{figure}

\begin{figure}
\includegraphics[width=0.49\textwidth]{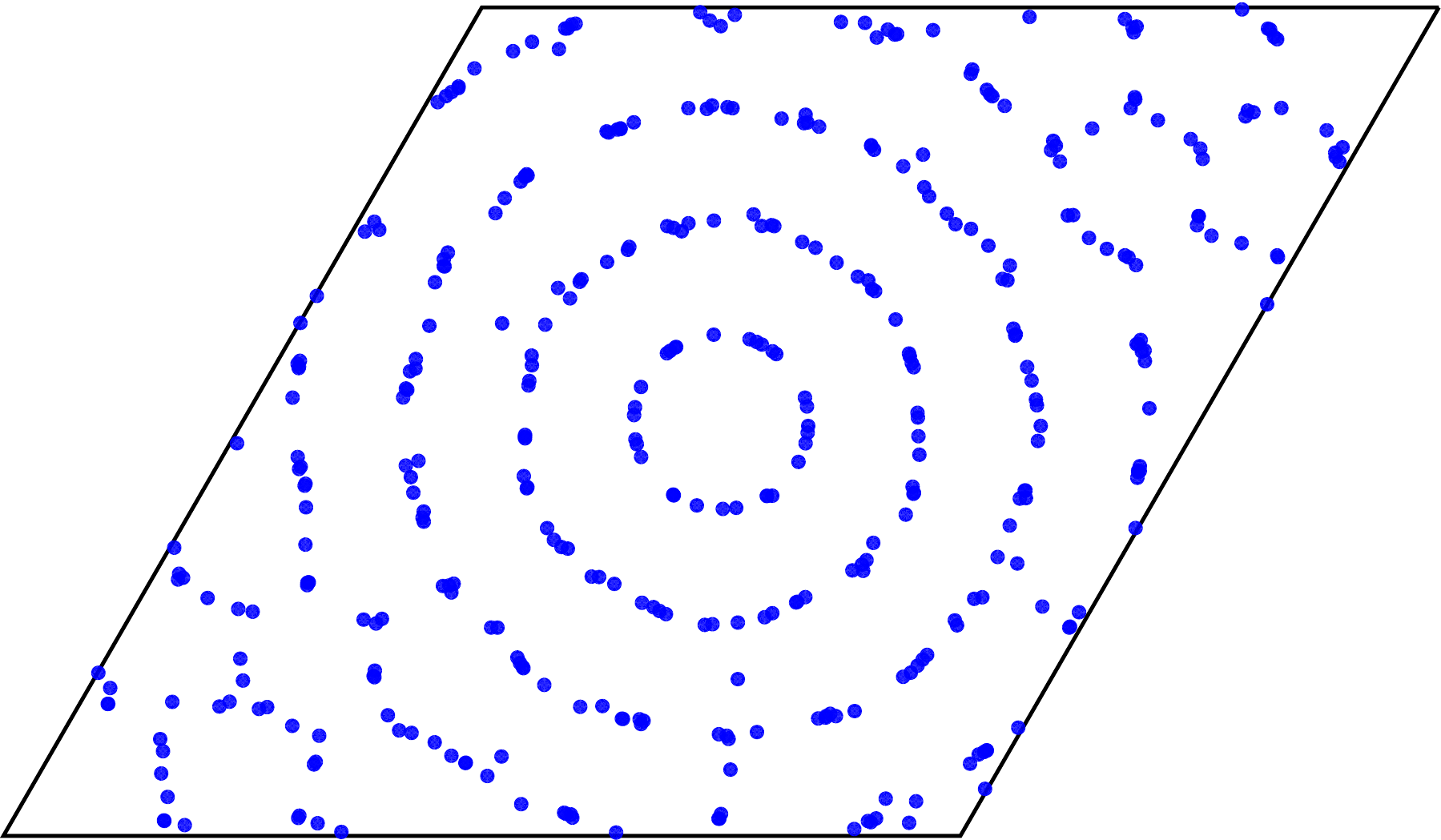}
\caption{A 2D biased stealthy configuration of $N=400$ particles obtained by energy minimization using the stealthy potential of $K=1$ and an external field of radius $R_f=4.58$ at $\chi=0.1$.}
\label{fig:2DExclusion}
\end{figure}

\begin{figure*}
\begin{tabular}{c c c c}
\includegraphics[width=0.3\textwidth]{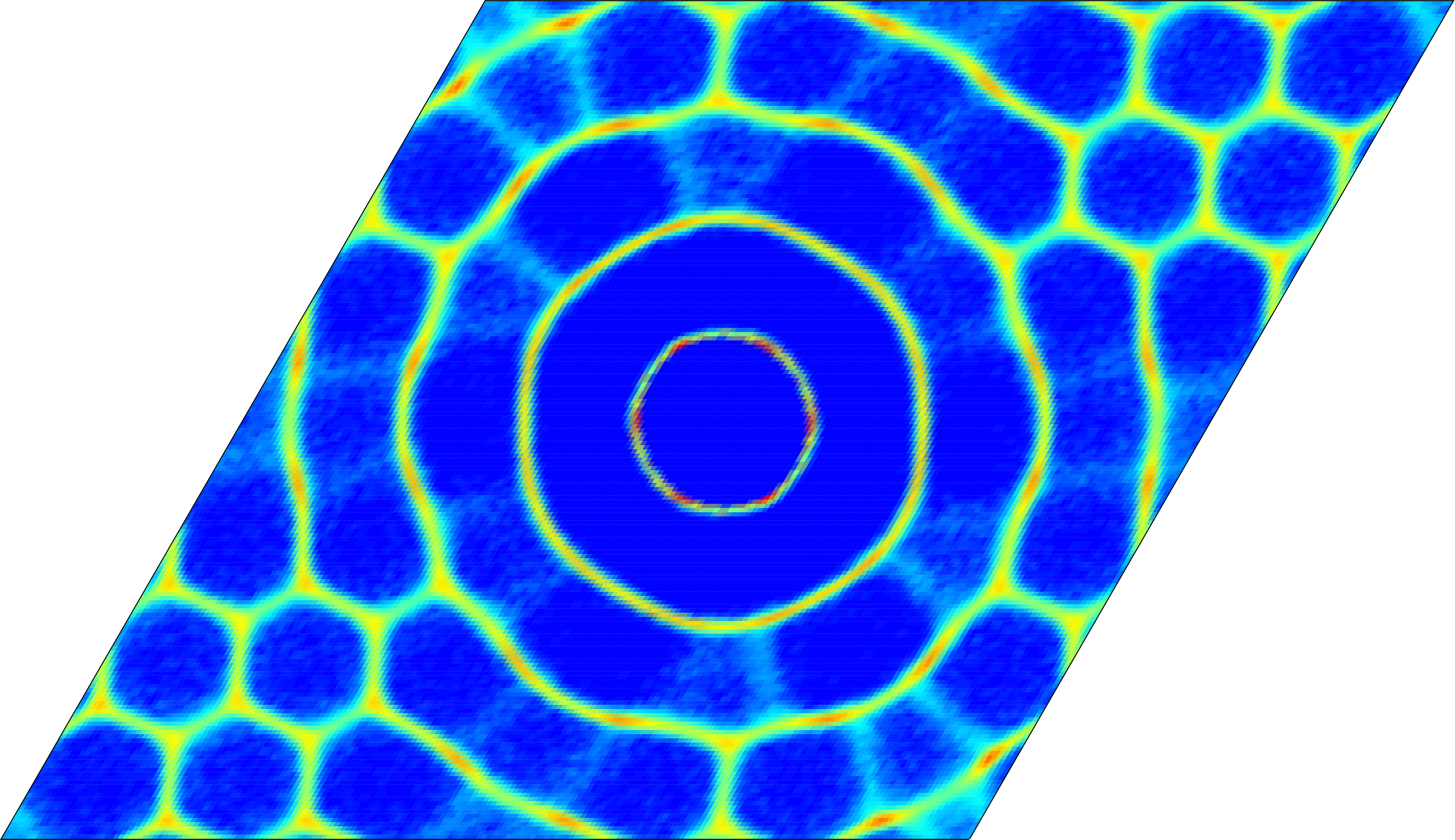}
&
\includegraphics[width=0.3\textwidth]{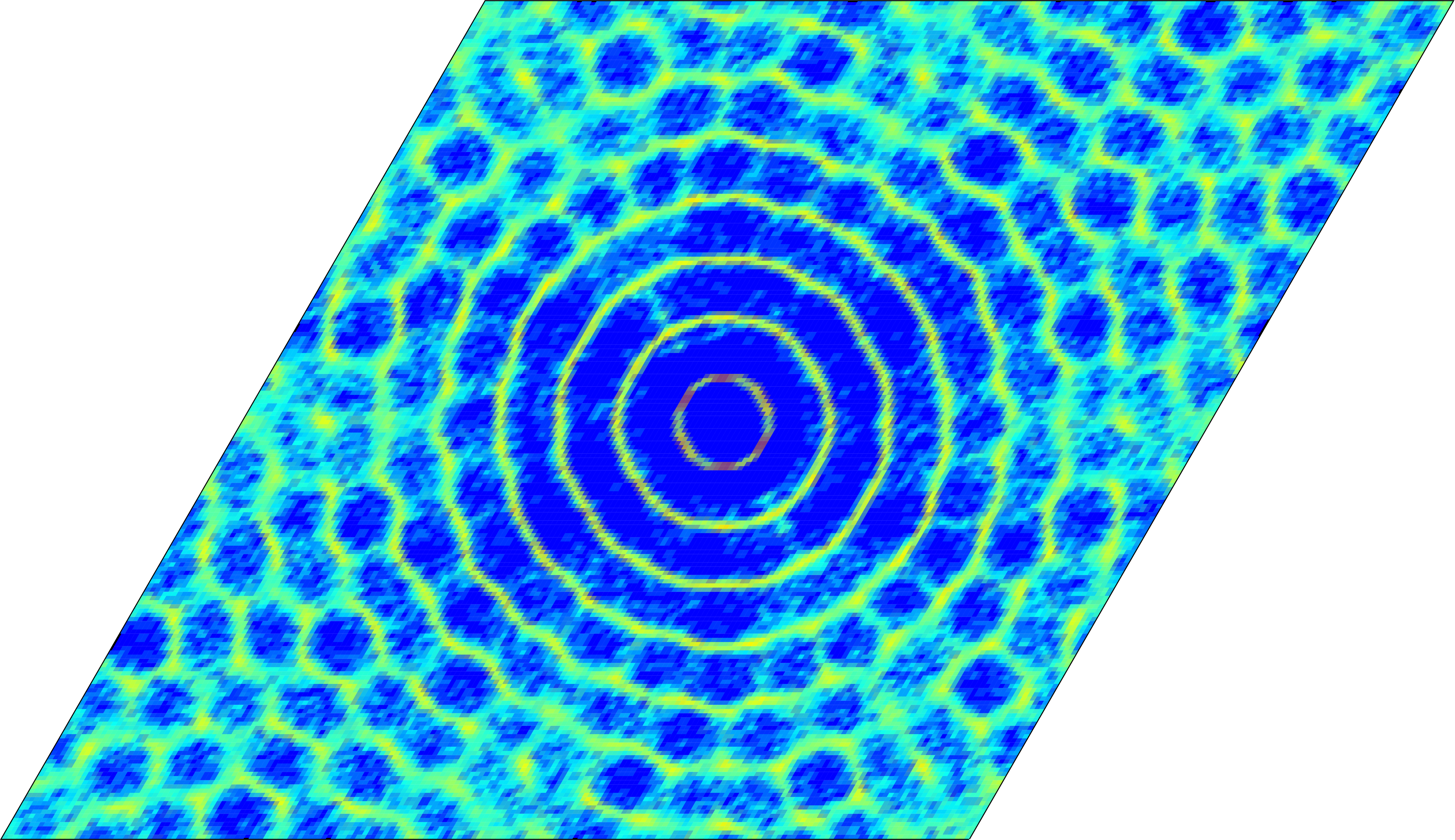}
&
\includegraphics[width=0.3\textwidth]{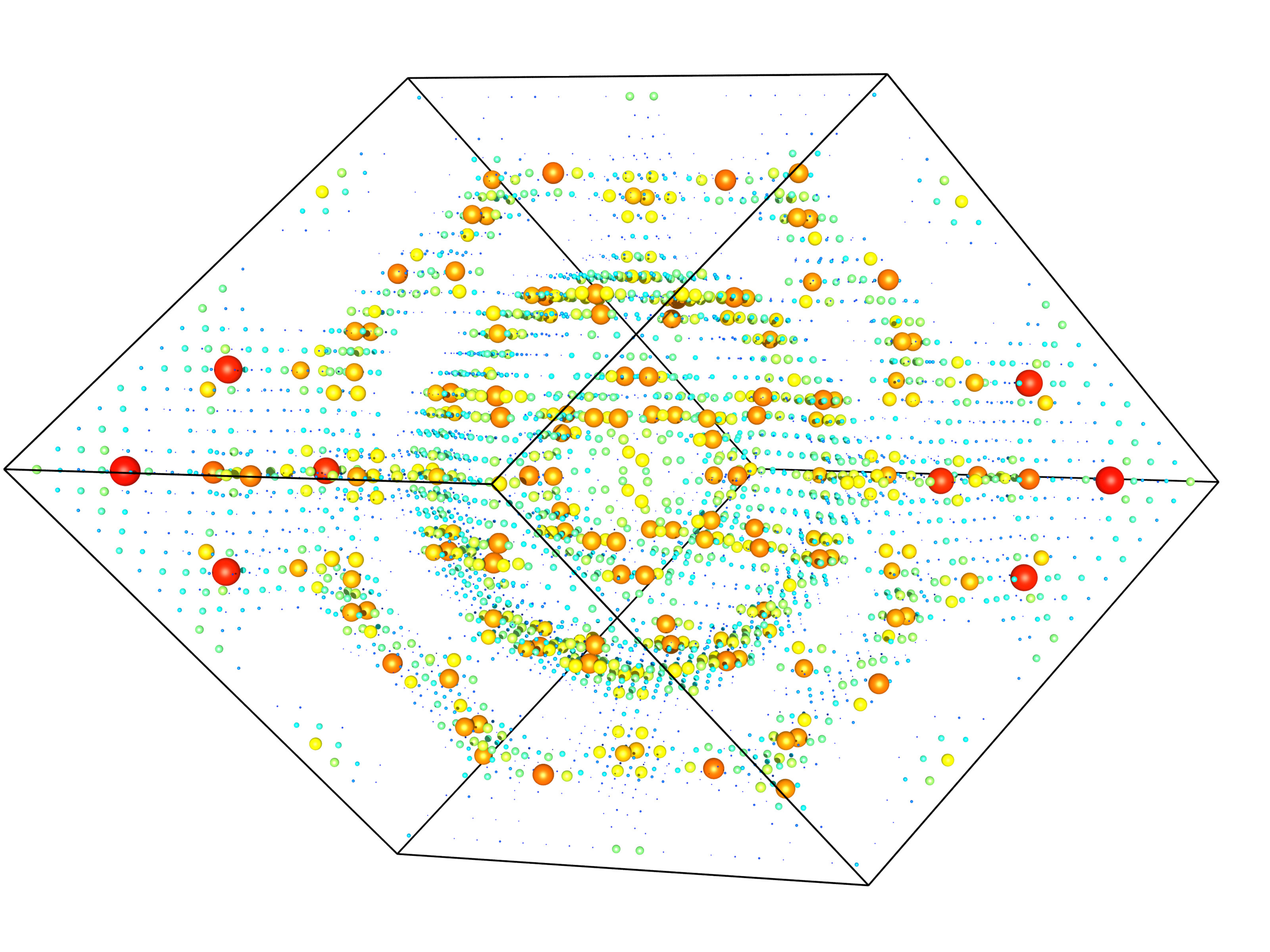}
&
\includegraphics[width=0.05\textwidth]{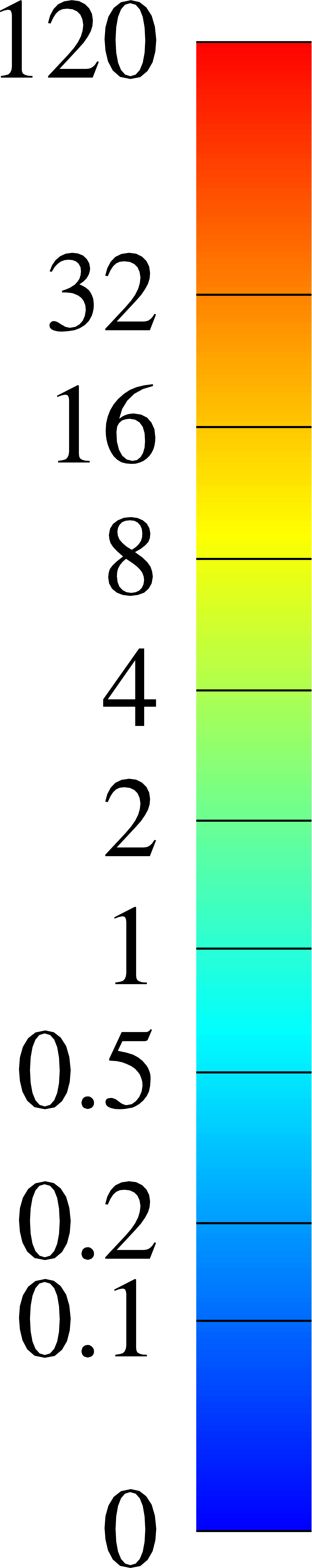} \\
A & B&C&\\

\end{tabular}
\caption{Numerically obtained $g_1(\mathbf r)$ for (A) $d=2$, $N=400$, $R_cK=4.58$, averaged over 3449 biased stealthy configurations; (B) $d=2$, $N=1600$, $R_cK=4.60$, averaged over 72 configurations; and (C) $d=3$, $N=400$, $R_cK=5.85$, averaged over 5174 configurations. The $\chi$ value is always 0.10.
In 3D, $g_1(\mathbf r)$ is represented by color-coded spheres with volumes proportional to $g_1(\mathbf r)$ at the spheres' location. Notice that there is a tendency for particles to concentrate in a lower-dimensional manifold.
}
\label{fig:g1}
\end{figure*}

We now move on to study maximum hole sizes in two and three dimensions. As we will see, these higher dimensions are computationally more challenging than 1D because the structures that maximize the hole size is not periodic. The $R_cK$ values found by the algorithm mentioned in Sec.~\ref{sec:Simulation} is presented in Fig.~\ref{fig:NumericalRc_K}. Similar to the 1D case, the dependence of $R_cK$ on $\chi$ or $N$ is weak and non-systematic. However, the 2D configurations, one of which is shown in Fig.~\ref{fig:2DExclusion}, exhibit a more complicated pattern, in which particles concentrate in a lower-dimensional manifold. Although this pattern is non-crystalline, it is still much more ordered than unbiased stealthy ground states at this $\chi$ value.\cite{zhang2015ground} To better reveal this pattern, we computed the one-body correlation function, $g_1(\mathbf r)$, of a 2D system of $\chi=0.1$ and $N=400$, shown in Fig.~\ref{fig:g1}A. The plot shows high-intensity concentric shells around the exclusion field (located at the center of the simulation box) and honeycomb network structures away from the exclusion field. Figure~\ref{fig:g1}B also shows $g_1(\mathbf r)$ of a larger 2D system, which exhibits the same pattern. Figure~\ref{fig:g1}C shows $g_1(\mathbf r)$ of a 3D system, which again has concentric shells around the exclusion field, but the structure away from the center is not obvious.

By pushing $R_cK$ to its numerical limit, we obtain periodic structures in 1D but non-periodic structures in 2D and 3D. Is it possible that this transition from periodic structures to non-periodic structures arises from increased numerical difficulties in higher dimensions? To eliminate this possibility, we analytically calculated $R_cK$ values for various 2D and 3D periodic structures for comparisons. In 2D, crystal structures achieve $R_cK=4.44$ but the system shown in Fig.~\ref{fig:g1} achieved $R_cK=4.6$; while in 3D crystal structures achieve $R_cK=5.44$ but the system shown in Fig.~\ref{fig:g1} achieved $R_cK=5.85$. Therefore, these non-periodic structures indeed have the largest known value of $R_cK$.

\begin{table}[h!]
\setlength{\tabcolsep}{12pt}
\caption{Maximum dimensionless hole size, $R_cK$, for various 2D crystalline structures.}
\begin{tabular}{c c}
\hline
Crystal & $R_cK$ \\ \hline
Square lattice & 4.44\\
Honeycomb crystal & 4.19\\
Triangular lattice & 4.19\\
Kagome crystal & 3.63\\
\hline
\end{tabular}
\label{tab:RcK_2D}
\end{table}

\begin{table}[h!]
\setlength{\tabcolsep}{12pt}
\caption{Maximum dimensionless hole size, $R_cK$, for various 3D crystalline structures.}
\begin{tabular}{c c}
\hline
Crystal & $R_cK$ \\ \hline
Face-centered cubic & 5.44\\
Simple cubic & 5.44\\
Hexagonal close packed & 5.13\\
Mean centered-cuboidal lattice \cite{conway1994lattices} & 5.03\\
Body-centered cubic & 4.97\\
Simple Hexagonal & 4.80\\
Diamond & 4.71\\
Pyrochlore crystal \cite{ramirez1999zero} & 4.51\\
\hline
\end{tabular}
\label{tab:RcK_3D}
\end{table}

It would be useful to analytically model these $g_1(\mathbf r)$ functions to find the maximum dimensionless hole size in the infinite-system-size limit. 
We will focus on the rings before considering the honeycomb-like structure away from the hole center. Comparing Fig.~\ref{fig:g1}A with Fig.~\ref{fig:g1}B, we see that increasing $N$ increases the number of rings. Therefore, we expect infinitely many rings in the infinite-system-size limit. 

\begin{figure}
\includegraphics[width=0.49\textwidth]{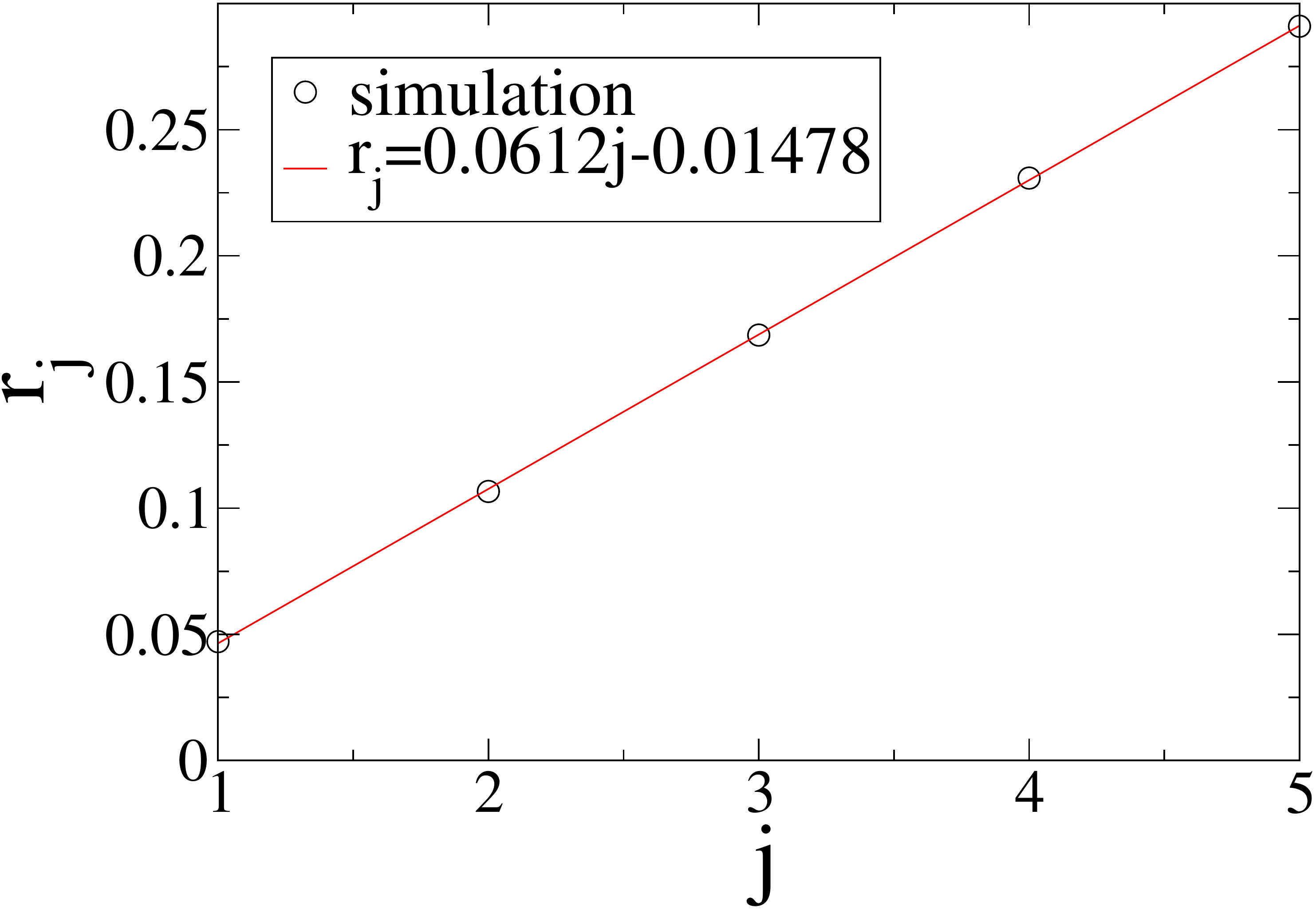}
\caption{The peak locations of $g_1(r)$ for a 2D biased stealthy system of $N=1600$ particles, at $\chi=0.10$, with an exclusion hole of dimensionless radius $R_cK=4.60$ at the origin, averaged over 72 configurations.}
\label{fig:g1peak}
\end{figure}

It is instructive to model an isotropic collection of concentric shells, for which we can write
\begin{equation}
g_1(\mathbf r)=\sum_{j=1}^\infty c_j \delta(|\mathbf r| - r_j),
\end{equation}
where $c_j$ is the intensity of the shells, $\delta$ is the Dirac delta function, and $r_j$ is the location of the shells. To determine $c_j$ and $r_j$, we computed the angular average of  $g_1(\mathbf r)$ shown in Fig.~\ref{fig:g1}B, and identified five peaks from it. As Fig.~\ref{fig:g1peak} shows, $r_j$ appears linear with $j$, for which linear regression produces $r_j=0.0612j-0.01478$. By rescaling the configuration, we can eliminate one fitting parameter and get $r_j=j-b$, where $b=0.242$.

To find $c_j$, we have computed the fraction of particles located on each ring, $p_j$. We find again $p_j$ is linear with $j$, with linear regression result $p_j=0.0275(j-0.242)\propto r_j$. Because $p_j$ is proportional to $r_j$, and is therefore proportional to the circumference of the rings, each ring has the same intensity. Neglecting a constant factor, we can then set $c_j=1$.

To summarize, numerical results suggest that in the infinite-system-size limit,
\begin{equation}
g_1(\mathbf r)\propto\sum_{j=1}^\infty \delta(|\mathbf r| - j+b),
\label{eq:g1}
\end{equation}
where constant $b$ is numerically measured as 0.242 in 2D. Note that this equation also applies to the 1D numerical result (an integer lattice of particle clusters) if we let $b=1/2$. The hole radius of this system is simply $R_c=1-b$, the radius of the first ring. After determining $R_c$, we should then ascertain $K$. Since $S(\mathbf k)$ is zero for all $\mathbf k$ such that $0<|\mathbf k|<K$, the collective coordinates ${\tilde \rho}(\mathbf k)=\sum_{j=1}^N \exp(-i \mathbf k \cdot \mathbf r_j)$ should also be zero. Thus, the Fourier transform of $g_1(\mathbf r)$, which we denote by ${\tilde g}_1(\mathbf k)$, should also be zero in this range. Fourier transforming Eq.~(\ref{eq:g1}) gives
\begin{equation}
{\tilde g}_1(\mathbf k)=\int_{\mathbf r} \exp(i\mathbf k \cdot \mathbf r) g_1(\mathbf r) \propto \sum_{j=1}^\infty \frac{(j-b)^{d/2}}{k^{d/2-1}}J_{d/2-1}[k(j-b)],
\label{eq:g1tilde}
\end{equation}
where $J_\nu$ is the Bessel function of order $\nu$. In Eq.~(\ref{eq:g1tilde}), letting $d=1$, 2, and 3 respectively yields
\begin{equation}
{\tilde g}_1(k)\propto \sum_{j=1}^\infty \cos[k(j-b)] \qquad \mbox{$(d=1)$},
\label{eq:tildeG1}
\end{equation}
\begin{equation}
{\tilde g}_1(k)\propto\sum_{j=1}^\infty (j-b)J_0[k(j-b)]\qquad \mbox{$(d=2)$},
\label{eq:tildeG2}
\end{equation}
and
\begin{equation}
{\tilde g}_1(k)\propto \sum_{j=1}^\infty (j-b)\cos[k(j-b)-\pi/2]\qquad \mbox{$(d=3)$}.
\label{eq:tildeG3}
\end{equation}
For large $x$, $J_0(x)$ is asymptotically $x^{-1/2}\cos(x-\pi/4)$. Substituting this into Eq.~(\ref{eq:tildeG2}) gives
\begin{equation}
{\tilde g}_1(k)\propto\sum_{j=1}^\infty \sqrt{\frac{j-b}{k}}\cos\left[k(j-b)-\frac{\pi}{4}\right]\quad \mbox{ $(d=2)$}.
\label{eq:tildeG22}
\end{equation}

We have already seen in the previous section that, the solution to maximizing $R_cK=(1-b)K$ in 1D is $b=1/2$ and $K=2\pi$. Comparing Eq.~(\ref{eq:tildeG22}) with Eq.~(\ref{eq:tildeG1}), in light of the numerical result $b \approx 0.242 \mbox{ $(d=2)$}$, suggests that $b=1/4$ in 2D. Somehow the $\pi/4$ phase factor in Eq.~(\ref{eq:tildeG22}) changes $b$ to $1/4$. If $K$ is still $2\pi$, then in 2D we have $R_cK=(1-b)K=3\pi/2\approx 4.71$, which is slightly above the numerically observed maximum dimensionless hole size $R_cK=4.65$. Similarly, in 3D, the $\pi/2$ phase factor in Eq.~(\ref{eq:tildeG3}) probably changes $b$ to $0$. If so, the maximum dimensionless hole size in 3D would be $R_cK=2\pi$. The difference between $2\pi$ and the numerically observed maximum $R_cK=5.86$ is nontrivial, but this can be explained by the increased numerical difficulty in 3D; for example, fewer concentric shells can be formed with the same number of particles in higher dimensions.

\section{Conclusions}
\label{sec:Conclusions}

In this paper, we have investigated the possibility of creating large holes in stealthy hyperuniform many-particle systems using numerical and analytical techniques. We demonstrated that hole sizes in such systems are bounded, first by examining the tail of $E_V(r)$ in unbiased ground states of stealthy potentials, and then by imposing radial exclusion fields to bias stealthy configurations toward ones with the largest possible holes. 
These results suggest that holes larger than a certain upper bound cannot exist in such systems. 
We then found that $R_cK$ is bounded from above by $\pi$, $3\pi/2$, and $2\pi$ in one, two, and three dimensions. A conjectured formula for the upper bound on the dimensionless hole size in $d$ dimensions is $(d+1)\pi/2$. An outstanding problem is a rigorous proof that stealthy infinite systems cannot tolerate holes of arbitrarily large sizes.

Our methods should be applicable to study the existence of arbitrarily large holes in other disordered many-particle systems. This is useful because maximum hole sizes and hole probabilities are related to several other important quantities, including the principal relaxation time $T_1$ associated with diffusion-controlled
reactions among traps. Specifically, consider a reactive chemical species that can diffuse in the void space between particles, and can be absorbed when it is within a certain distance to any particle. The fraction of such species, released at time $t=0$, that is not absorbed at time $t$ (in other words, the survival probability of the molecules of such species), can be expanded as a series of exponential functions \cite{torquato1991diffusion}
\begin{equation}
p(t) = \sum_{n=1}^\infty I_n \exp(-t/T_n),
\end{equation}
where $I_n$ are coefficients and $T_n$ are relaxation times. The largest relaxation time $T_1$ is called the ``principal relaxation time.'' The relaxation times can be measured directly by NMR experiments, in which proton magnetization decays at the phase boundary.\cite{straley1987magnetic, banavar1987magnetic, mitra1992effects} It has been demonstrated that $T_1$ is determined by the largest holes in the configurations, and is therefore divergent if arbitrarily large holes can occur.\cite{torquato1991diffusion}  Indeed, for a reactive species in equilibrium hard-sphere systems, the large-$t$ behavior of its survival probability is actually $p(t)\sim \exp[-t^{3/5}]$ in three dimensions.\cite{torquato1991diffusion} It is noteworthy that stealthy trap model systems that prohibit arbitrarily large holes would have finite $T_1$ values.

It is noteworthy that the maximum hole size of a solvent is also related to the largest solute particles that it can dissolve. In a solvent with a finite value of $R_c$, particles with exclusion radius larger than $R_c$ would create intolerably large holes, and would therefore not dissolve. Solute particles smaller than $R_c$ would dissolve in a strictly stealthy solvent, but the effective interactions between them deserve future research. Would particles larger than $R_c/2$ refuse to touch each other in order to avoid combining the holes they create? Also, if the solute particles are only slightly smaller than $R_c$, solvent particles should be concentrated in concentric-shell regions around the solute particles. Could the interference between these concentric shells induce very complicated effective interactions?

\appendix
\section{Expected $R_c$ for a finite number of finite-sized Poisson configurations}
\label{app:PoissonRc}
Although there is no theoretical limit on the hole radii in Poisson configurations (ideal gas), one still expects to find a finite $R_c$ if one only studies a finite number of finite sized configurations. If one studies a total of $N_c$ configurations of $N$ particles, one expects to see roughly $N_cN$ uncorrelated holes. Of these $N_cN$ holes, one expects to find the largest hole once. Therefore
\begin{equation}
E_V(R_c)=\exp[-\rho v_1(R_c)]=\frac{1}{N_cN}.
\label{eq:A1}
\end{equation}
This equation predicts the largest hole size, $R_c$, as a function of $\rho$, $N_c$, and $N$. To find $R_c$ presented in Fig.~\ref{fig:Stealthy_Rc}, notice that for stealthy systems of a given $\chi$ and $K$, $\rho$ is given in Eq.~(\ref{eq:rhochi}). Substituting Eq.~(\ref{eq:rhochi}) into Eq.~(\ref{eq:A1}) yields
\begin{equation}
\frac{v_1(R_c)v_1(K)}{2d\chi(2\pi)^d}=\ln(N_cN).
\end{equation}
Here we use $K=1$, $N_c=20000$, and $N=500$ to be consistent with stealthy results.

\footnotesize{
\providecommand*{\mcitethebibliography}{\thebibliography}
\csname @ifundefined\endcsname{endmcitethebibliography}
{\let\endmcitethebibliography\endthebibliography}{}

}


\begin{mcitethebibliography}{43}
\providecommand*{\natexlab}[1]{#1}
\providecommand*{\mciteSetBstSublistMode}[1]{}
\providecommand*{\mciteSetBstMaxWidthForm}[2]{}
\providecommand*{\mciteBstWouldAddEndPuncttrue}
  {\def\EndOfBibitem{\unskip.}}
\providecommand*{\mciteBstWouldAddEndPunctfalse}
  {\let\EndOfBibitem\relax}
\providecommand*{\mciteSetBstMidEndSepPunct}[3]{}
\providecommand*{\mciteSetBstSublistLabelBeginEnd}[3]{}
\providecommand*{\EndOfBibitem}{}
\mciteSetBstSublistMode{f}
\mciteSetBstMaxWidthForm{subitem}
{(\emph{\alph{mcitesubitemcount}})}
\mciteSetBstSublistLabelBeginEnd{\mcitemaxwidthsubitemform\space}
{\relax}{\relax}

\bibitem[Yarnell \emph{et~al.}(1973)Yarnell, Katz, Wenzel, and
  Koenig]{yarnell1973structure}
J.~Yarnell, M.~Katz, R.~G. Wenzel and S.~Koenig, \emph{Phys. Rev. A}, 1973,
  \textbf{7}, 2130\relax
\mciteBstWouldAddEndPuncttrue
\mciteSetBstMidEndSepPunct{\mcitedefaultmidpunct}
{\mcitedefaultendpunct}{\mcitedefaultseppunct}\relax
\EndOfBibitem
\bibitem[Ortiz and Ballone(1994)]{ortiz1994correlation}
G.~Ortiz and P.~Ballone, \emph{Phys. Rev. B}, 1994, \textbf{50}, 1391\relax
\mciteBstWouldAddEndPuncttrue
\mciteSetBstMidEndSepPunct{\mcitedefaultmidpunct}
{\mcitedefaultendpunct}{\mcitedefaultseppunct}\relax
\EndOfBibitem
\bibitem[Chandler(1987)]{chandler1987introduction}
D.~Chandler, \emph{Introduction to Modern Statistical Mechanics}, Oxford
  University Press, 1987\relax
\mciteBstWouldAddEndPuncttrue
\mciteSetBstMidEndSepPunct{\mcitedefaultmidpunct}
{\mcitedefaultendpunct}{\mcitedefaultseppunct}\relax
\EndOfBibitem
\bibitem[Filipponi \emph{et~al.}(1990)Filipponi, Di~Cicco, Benfatto, and
  Natoli]{filipponi1990three}
A.~Filipponi, A.~Di~Cicco, M.~Benfatto and C.~Natoli, \emph{Europhys. Lett.},
  1990, \textbf{13}, 319\relax
\mciteBstWouldAddEndPuncttrue
\mciteSetBstMidEndSepPunct{\mcitedefaultmidpunct}
{\mcitedefaultendpunct}{\mcitedefaultseppunct}\relax
\EndOfBibitem
\bibitem[Torquato \emph{et~al.}(1990)Torquato, Lu, and
  Rubinstein]{torquato1990nearest}
S.~Torquato, B.~Lu and J.~Rubinstein, \emph{Phys. Rev. A}, 1990, \textbf{41},
  2059\relax
\mciteBstWouldAddEndPuncttrue
\mciteSetBstMidEndSepPunct{\mcitedefaultmidpunct}
{\mcitedefaultendpunct}{\mcitedefaultseppunct}\relax
\EndOfBibitem
\bibitem[Torquato(1995)]{torquato1995nearest}
S.~Torquato, \emph{Phys. Rev. E}, 1995, \textbf{51}, 3170\relax
\mciteBstWouldAddEndPuncttrue
\mciteSetBstMidEndSepPunct{\mcitedefaultmidpunct}
{\mcitedefaultendpunct}{\mcitedefaultseppunct}\relax
\EndOfBibitem
\bibitem[Starr \emph{et~al.}(2002)Starr, Sastry, Douglas, and
  Glotzer]{starr2002we}
F.~W. Starr, S.~Sastry, J.~F. Douglas and S.~C. Glotzer, \emph{Phys. Rev.
  Lett.}, 2002, \textbf{89}, 125501\relax
\mciteBstWouldAddEndPuncttrue
\mciteSetBstMidEndSepPunct{\mcitedefaultmidpunct}
{\mcitedefaultendpunct}{\mcitedefaultseppunct}\relax
\EndOfBibitem
\bibitem[S.~Kumar and Kumaran(2005)]{senthil2005voronoi}
V.~S.~Kumar and V.~Kumaran, \emph{J. Chem. Phys.}, 2005, \textbf{123},
  114501\relax
\mciteBstWouldAddEndPuncttrue
\mciteSetBstMidEndSepPunct{\mcitedefaultmidpunct}
{\mcitedefaultendpunct}{\mcitedefaultseppunct}\relax
\EndOfBibitem
\bibitem[Hentschel \emph{et~al.}(2007)Hentschel, Ilyin, Makedonska, Procaccia,
  and Schupper]{hentschel2007statistical}
H.~Hentschel, V.~Ilyin, N.~Makedonska, I.~Procaccia and N.~Schupper,
  \emph{Phys. Rev. E}, 2007, \textbf{75}, 050404\relax
\mciteBstWouldAddEndPuncttrue
\mciteSetBstMidEndSepPunct{\mcitedefaultmidpunct}
{\mcitedefaultendpunct}{\mcitedefaultseppunct}\relax
\EndOfBibitem
\bibitem[Slotterback \emph{et~al.}(2008)Slotterback, Toiya, Goff, Douglas, and
  Losert]{slotterback2008correlation}
S.~Slotterback, M.~Toiya, L.~Goff, J.~F. Douglas and W.~Losert, \emph{Phys.
  Rev. Lett.}, 2008, \textbf{101}, 258001\relax
\mciteBstWouldAddEndPuncttrue
\mciteSetBstMidEndSepPunct{\mcitedefaultmidpunct}
{\mcitedefaultendpunct}{\mcitedefaultseppunct}\relax
\EndOfBibitem
\bibitem[Schr{\"o}der-Turk \emph{et~al.}(2011)Schr{\"o}der-Turk, Mickel,
  Kapfer, Klatt, Schaller, Hoffmann, Kleppmann, Armstrong, Inayat,
  Hug,\emph{et~al.}]{schroder2011minkowski}
G.~Schr{\"o}der-Turk, W.~Mickel, S.~Kapfer, M.~Klatt, F.~Schaller, M.~Hoffmann,
  N.~Kleppmann, P.~Armstrong, A.~Inayat, D.~Hug \emph{et~al.}, \emph{Adv.
  Mater.}, 2011, \textbf{23}, 2535--2553\relax
\mciteBstWouldAddEndPuncttrue
\mciteSetBstMidEndSepPunct{\mcitedefaultmidpunct}
{\mcitedefaultendpunct}{\mcitedefaultseppunct}\relax
\EndOfBibitem
\bibitem[Ma(2015)]{ma2015tuning}
E.~Ma, \emph{Nat. Mater.}, 2015, \textbf{14}, 547--552\relax
\mciteBstWouldAddEndPuncttrue
\mciteSetBstMidEndSepPunct{\mcitedefaultmidpunct}
{\mcitedefaultendpunct}{\mcitedefaultseppunct}\relax
\EndOfBibitem
\bibitem[Torquato(2001)]{torquato2001random}
S.~Torquato, \emph{Random heterogeneous materials: microstructure and
  macroscopic properties}, Springer Science \& Business Media, 2001,
  vol.~16\relax
\mciteBstWouldAddEndPuncttrue
\mciteSetBstMidEndSepPunct{\mcitedefaultmidpunct}
{\mcitedefaultendpunct}{\mcitedefaultseppunct}\relax
\EndOfBibitem
\bibitem[Torquato(2010)]{torquato2010reformulation}
S.~Torquato, \emph{Phys. Rev. E}, 2010, \textbf{82}, 056109\relax
\mciteBstWouldAddEndPuncttrue
\mciteSetBstMidEndSepPunct{\mcitedefaultmidpunct}
{\mcitedefaultendpunct}{\mcitedefaultseppunct}\relax
\EndOfBibitem
\bibitem[Uche \emph{et~al.}(2004)Uche, Stillinger, and
  Torquato]{uche2004constraints}
O.~U. Uche, F.~H. Stillinger and S.~Torquato, \emph{Phys. Rev. E}, 2004,
  \textbf{70}, 046122\relax
\mciteBstWouldAddEndPuncttrue
\mciteSetBstMidEndSepPunct{\mcitedefaultmidpunct}
{\mcitedefaultendpunct}{\mcitedefaultseppunct}\relax
\EndOfBibitem
\bibitem[Batten \emph{et~al.}(2008)Batten, Stillinger, and
  Torquato]{batten2008classical}
R.~D. Batten, F.~H. Stillinger and S.~Torquato, \emph{J. Appl. Phys.}, 2008,
  \textbf{104}, 033504--033504\relax
\mciteBstWouldAddEndPuncttrue
\mciteSetBstMidEndSepPunct{\mcitedefaultmidpunct}
{\mcitedefaultendpunct}{\mcitedefaultseppunct}\relax
\EndOfBibitem
\bibitem[Torquato \emph{et~al.}(2015)Torquato, Zhang, and
  Stillinger]{torquato2015ensemble}
S.~Torquato, G.~Zhang and F.~Stillinger, \emph{Phys. Rev. X}, 2015, \textbf{5},
  021020\relax
\mciteBstWouldAddEndPuncttrue
\mciteSetBstMidEndSepPunct{\mcitedefaultmidpunct}
{\mcitedefaultendpunct}{\mcitedefaultseppunct}\relax
\EndOfBibitem
\bibitem[Zhang \emph{et~al.}(2015)Zhang, Stillinger, and
  Torquato]{zhang2015ground}
G.~Zhang, F.~H. Stillinger and S.~Torquato, \emph{Phys. Rev. E}, 2015,
  \textbf{92}, 022119\relax
\mciteBstWouldAddEndPuncttrue
\mciteSetBstMidEndSepPunct{\mcitedefaultmidpunct}
{\mcitedefaultendpunct}{\mcitedefaultseppunct}\relax
\EndOfBibitem
\bibitem[Zhang \emph{et~al.}(2015)Zhang, Stillinger, and
  Torquato]{zhang2015ground2}
G.~Zhang, F.~H. Stillinger and S.~Torquato, \emph{Phys. Rev. E}, 2015,
  \textbf{92}, 022120\relax
\mciteBstWouldAddEndPuncttrue
\mciteSetBstMidEndSepPunct{\mcitedefaultmidpunct}
{\mcitedefaultendpunct}{\mcitedefaultseppunct}\relax
\EndOfBibitem
\bibitem[Batten \emph{et~al.}(2009)Batten, Stillinger, and
  Torquato]{batten2009novel}
R.~D. Batten, F.~H. Stillinger and S.~Torquato, \emph{Phys. Rev. Lett.}, 2009,
  \textbf{103}, 050602\relax
\mciteBstWouldAddEndPuncttrue
\mciteSetBstMidEndSepPunct{\mcitedefaultmidpunct}
{\mcitedefaultendpunct}{\mcitedefaultseppunct}\relax
\EndOfBibitem
\bibitem[Florescu \emph{et~al.}(2009)Florescu, Torquato, and
  Steinhardt]{florescu2009designer}
M.~Florescu, S.~Torquato and P.~J. Steinhardt, \emph{Proc. Natl. Acad. Sci.},
  2009, \textbf{106}, 20658--20663\relax
\mciteBstWouldAddEndPuncttrue
\mciteSetBstMidEndSepPunct{\mcitedefaultmidpunct}
{\mcitedefaultendpunct}{\mcitedefaultseppunct}\relax
\EndOfBibitem
\bibitem[Florescu \emph{et~al.}(2013)Florescu, Steinhardt, and
  Torquato]{florescu2013optical}
M.~Florescu, P.~J. Steinhardt and S.~Torquato, \emph{Phys. Rev. B}, 2013,
  \textbf{87}, 165116\relax
\mciteBstWouldAddEndPuncttrue
\mciteSetBstMidEndSepPunct{\mcitedefaultmidpunct}
{\mcitedefaultendpunct}{\mcitedefaultseppunct}\relax
\EndOfBibitem
\bibitem[Man \emph{et~al.}(2013)Man, Florescu, Williamson, He, Hashemizad,
  Leung, Liner, Torquato, Chaikin, and Steinhardt]{man2013isotropic}
W.~Man, M.~Florescu, E.~P. Williamson, Y.~He, S.~R. Hashemizad, B.~Y.~C. Leung,
  D.~R. Liner, S.~Torquato, P.~M. Chaikin and P.~J. Steinhardt, \emph{Proc.
  Natl. Acad. Sci.}, 2013, \textbf{110}, 15886--15891\relax
\mciteBstWouldAddEndPuncttrue
\mciteSetBstMidEndSepPunct{\mcitedefaultmidpunct}
{\mcitedefaultendpunct}{\mcitedefaultseppunct}\relax
\EndOfBibitem
\bibitem[Leseur \emph{et~al.}(2016)Leseur, Pierrat, and
  Carminati]{leseur2016high}
O.~Leseur, R.~Pierrat and R.~Carminati, \emph{Optica}, 2016, \textbf{3},
  763\relax
\mciteBstWouldAddEndPuncttrue
\mciteSetBstMidEndSepPunct{\mcitedefaultmidpunct}
{\mcitedefaultendpunct}{\mcitedefaultseppunct}\relax
\EndOfBibitem
\bibitem[Zhang \emph{et~al.}(2016)Zhang, Stillinger, and
  Torquato]{zhang2016transport}
G.~Zhang, F.~Stillinger and S.~Torquato, \emph{J. Chem. Phys.}, 2016,
  \textbf{145}, 244109\relax
\mciteBstWouldAddEndPuncttrue
\mciteSetBstMidEndSepPunct{\mcitedefaultmidpunct}
{\mcitedefaultendpunct}{\mcitedefaultseppunct}\relax
\EndOfBibitem
\bibitem[Conway and Sloane(1998)]{conway2013sphere}
J.~H. Conway and N.~J.~A. Sloane, \emph{Sphere packings, lattices and groups},
  Springer Science \& Business Media, 1998\relax
\mciteBstWouldAddEndPuncttrue
\mciteSetBstMidEndSepPunct{\mcitedefaultmidpunct}
{\mcitedefaultendpunct}{\mcitedefaultseppunct}\relax
\EndOfBibitem
\bibitem[Atkinson \emph{et~al.}(2016)Atkinson, Stillinger, and
  Torquato]{atkinson2016static}
S.~Atkinson, F.~H. Stillinger and S.~Torquato, \emph{Phys. Rev. E}, 2016,
  \textbf{94}, 032902\relax
\mciteBstWouldAddEndPuncttrue
\mciteSetBstMidEndSepPunct{\mcitedefaultmidpunct}
{\mcitedefaultendpunct}{\mcitedefaultseppunct}\relax
\EndOfBibitem
\bibitem[Rintoul \emph{et~al.}(1996)Rintoul, Torquato, and
  Tarjus]{rintoul1996nearest}
M.~Rintoul, S.~Torquato and G.~Tarjus, \emph{Phys. Rev. E}, 1996, \textbf{53},
  450\relax
\mciteBstWouldAddEndPuncttrue
\mciteSetBstMidEndSepPunct{\mcitedefaultmidpunct}
{\mcitedefaultendpunct}{\mcitedefaultseppunct}\relax
\EndOfBibitem
\bibitem[Torquato and Stillinger(2003)]{torquato2003local}
S.~Torquato and F.~H. Stillinger, \emph{Phys. Rev. E}, 2003, \textbf{68},
  041113\relax
\mciteBstWouldAddEndPuncttrue
\mciteSetBstMidEndSepPunct{\mcitedefaultmidpunct}
{\mcitedefaultendpunct}{\mcitedefaultseppunct}\relax
\EndOfBibitem
\bibitem[Torquato \emph{et~al.}(2008)Torquato, Scardicchio, and
  Zachary]{torquato2008point}
S.~Torquato, A.~Scardicchio and C.~E. Zachary, \emph{J. Stat. Mech. Theor.
  Exp.}, 2008, \textbf{2008}, P11019\relax
\mciteBstWouldAddEndPuncttrue
\mciteSetBstMidEndSepPunct{\mcitedefaultmidpunct}
{\mcitedefaultendpunct}{\mcitedefaultseppunct}\relax
\EndOfBibitem
\bibitem[Hough \emph{et~al.}(2009)Hough, Krishnapur, Peres, and
  Vir{\'a}g]{Hough09}
J.~B. Hough, M.~Krishnapur, Y.~Peres and B.~Vir{\'a}g, \emph{Zeros of Gaussian
  analytic functions and determinantal point processes}, American Mathematical
  Society, Providence, RI, 2009, vol.~51\relax
\mciteBstWouldAddEndPuncttrue
\mciteSetBstMidEndSepPunct{\mcitedefaultmidpunct}
{\mcitedefaultendpunct}{\mcitedefaultseppunct}\relax
\EndOfBibitem
\bibitem[Ghosh and Nishry(2016)]{ghosh2016gaussian}
S.~Ghosh and A.~Nishry, \emph{arXiv preprint arXiv:1609.00084}, 2016\relax
\mciteBstWouldAddEndPuncttrue
\mciteSetBstMidEndSepPunct{\mcitedefaultmidpunct}
{\mcitedefaultendpunct}{\mcitedefaultseppunct}\relax
\EndOfBibitem
\bibitem[Chaikin and Lubensky(2000)]{chaikin2000principles}
P.~Chaikin and T.~Lubensky, \emph{Principles of Condensed Matter Physics},
  Cambridge University Press, 2000\relax
\mciteBstWouldAddEndPuncttrue
\mciteSetBstMidEndSepPunct{\mcitedefaultmidpunct}
{\mcitedefaultendpunct}{\mcitedefaultseppunct}\relax
\EndOfBibitem
\bibitem[Zhang and Torquato(2013)]{zhang2013precise}
G.~Zhang and S.~Torquato, \emph{Phys. Rev. E}, 2013, \textbf{88}, 053312\relax
\mciteBstWouldAddEndPuncttrue
\mciteSetBstMidEndSepPunct{\mcitedefaultmidpunct}
{\mcitedefaultendpunct}{\mcitedefaultseppunct}\relax
\EndOfBibitem
\bibitem[Nocedal(1980)]{nocedal1980updating}
J.~Nocedal, \emph{Math. Comp.}, 1980, \textbf{35}, 773--782\relax
\mciteBstWouldAddEndPuncttrue
\mciteSetBstMidEndSepPunct{\mcitedefaultmidpunct}
{\mcitedefaultendpunct}{\mcitedefaultseppunct}\relax
\EndOfBibitem
\bibitem[Liu and Nocedal(1989)]{liu1989limited}
D.~C. Liu and J.~Nocedal, \emph{Math. Programming}, 1989, \textbf{45},
  503--528\relax
\mciteBstWouldAddEndPuncttrue
\mciteSetBstMidEndSepPunct{\mcitedefaultmidpunct}
{\mcitedefaultendpunct}{\mcitedefaultseppunct}\relax
\EndOfBibitem
\bibitem[Johnson()]{nlopt}
S.~G. Johnson, \emph{The NLopt nonlinear-optimization package},
  http://ab-initio.mit.edu/nlopt\relax
\mciteBstWouldAddEndPuncttrue
\mciteSetBstMidEndSepPunct{\mcitedefaultmidpunct}
{\mcitedefaultendpunct}{\mcitedefaultseppunct}\relax
\EndOfBibitem
\bibitem[Conway and Sloane(1994)]{conway1994lattices}
J.~Conway and N.~Sloane, \emph{J. Number Theor.}, 1994, \textbf{48},
  373--382\relax
\mciteBstWouldAddEndPuncttrue
\mciteSetBstMidEndSepPunct{\mcitedefaultmidpunct}
{\mcitedefaultendpunct}{\mcitedefaultseppunct}\relax
\EndOfBibitem
\bibitem[Ramirez \emph{et~al.}(1999)Ramirez, Hayashi, Cava, Siddharthan, and
  Shastry]{ramirez1999zero}
A.~P. Ramirez, A.~Hayashi, R.~Cava, R.~Siddharthan and B.~Shastry,
  \emph{Nature}, 1999, \textbf{399}, 333--335\relax
\mciteBstWouldAddEndPuncttrue
\mciteSetBstMidEndSepPunct{\mcitedefaultmidpunct}
{\mcitedefaultendpunct}{\mcitedefaultseppunct}\relax
\EndOfBibitem
\bibitem[Torquato and Avellaneda(1991)]{torquato1991diffusion}
S.~Torquato and M.~Avellaneda, \emph{J. Chem. Phys.}, 1991, \textbf{95},
  6477--6489\relax
\mciteBstWouldAddEndPuncttrue
\mciteSetBstMidEndSepPunct{\mcitedefaultmidpunct}
{\mcitedefaultendpunct}{\mcitedefaultseppunct}\relax
\EndOfBibitem
\bibitem[Straley \emph{et~al.}(1987)Straley, Matteson, Feng, Schwartz, Kenyon,
  and Banavar]{straley1987magnetic}
C.~Straley, A.~Matteson, S.~Feng, L.~M. Schwartz, W.~E. Kenyon and J.~R.
  Banavar, \emph{Appl. Phys. Lett.}, 1987, \textbf{51}, 1146--1148\relax
\mciteBstWouldAddEndPuncttrue
\mciteSetBstMidEndSepPunct{\mcitedefaultmidpunct}
{\mcitedefaultendpunct}{\mcitedefaultseppunct}\relax
\EndOfBibitem
\bibitem[Banavar and Schwartz(1987)]{banavar1987magnetic}
J.~R. Banavar and L.~M. Schwartz, \emph{Phys. Rev. Lett.}, 1987, \textbf{58},
  1411\relax
\mciteBstWouldAddEndPuncttrue
\mciteSetBstMidEndSepPunct{\mcitedefaultmidpunct}
{\mcitedefaultendpunct}{\mcitedefaultseppunct}\relax
\EndOfBibitem
\bibitem[Mitra and Sen(1992)]{mitra1992effects}
P.~P. Mitra and P.~N. Sen, \emph{Phys. Rev. B}, 1992, \textbf{45}, 143\relax
\mciteBstWouldAddEndPuncttrue
\mciteSetBstMidEndSepPunct{\mcitedefaultmidpunct}
{\mcitedefaultendpunct}{\mcitedefaultseppunct}\relax
\EndOfBibitem
\end{mcitethebibliography}
\end{document}